\documentclass[editor]{aa}
\usepackage{newtxtext,newtxmath}
\usepackage[utf8]{inputenc}
\usepackage{ae,aecompl}
\usepackage{graphicx}
\usepackage{amsmath,amssymb}
\usepackage{subfig}
\usepackage{natbib}
\usepackage{subfiles}
\usepackage{siunitx}
\usepackage{pdflscape}
\usepackage[breaklinks,colorlinks,citecolor=blue]{hyperref}
\usepackage{cleveref}
\usepackage{rotating}
\usepackage{booktabs}
\usepackage{float}
\usepackage{placeins}
\usepackage{xcolor}

\begin{document}

\newcommand{\Rvis}{R$_{550}$}
\newcommand{\tauvis}{$\tau_{550}$}
\newcommand{\AV}{A$_{\rm V}$}
\newcommand{\micron}{$\mu m$}

\newcommand\cfm[1]{{\color{red}CFM:\bf#1}}
\newcommand\mb[1]{{\color{magenta}\bf#1}}
\newcommand\rb[1]{{\color{blue}\bf#1}}

\newcommand{\LiI}{\ion{Li}{i}}
\newcommand{\CaI}{\ion{Ca}{i}}

\title{The dance of dust: Investigating young stellar object dipper variability \thanks{Based on observations collected at the European Southern Observatory under ESO programmes 0105.C-0513(A),097.C-0378(A), 0101.C-0866(A), and 0103.C-0887(B).}}
\author{Empey, A.\inst{1} \and Garcia Lopez, R.\inst{1} 
\and Natta, A.\inst{1,2} \and Manara, C.F.\inst{3} \and Benisty, M. \inst{4} \and Claes, R. \inst{3} \and McGinnis, P. \inst{5}}
\institute{
University College Dublin (UCD), Department of Physics, Belfield, Dublin 4, Ireland 
\and Dublin Institute for Advanced Studies, Astronomy \& Astrophysics Section, 
31 Fitzwilliam Place, Dublin 2, D02 XF86, Ireland 
\and European Southern Observatory, Karl-Schwarzschild-Strasse 2, 85748 Garching bei M\"unchen, Germany
\and Max-Planck Institute for Astronomy(MPIA), Königstuhl 17, 69117 Heidelberg, Germany
\and Independent Researcher}
\date{Received 1st August 2025; Accepted 20th October 2025}

 \abstract
   {The dipper subclass of young stellar objects (YSOs) are characterised by frequent dips in their light curves. Irregular dippers do not show periodic signatures and have dips accounting for significant proportions of their photospheric flux. Given the short timescales on which these dips occur, their driving mechanisms are linked to the inner circumstellar disc dynamics.}
   {We present the first multi-epoch analysis of 16 irregular dippers observed with X-Shooter. Investigating the properties of their dips, and in particular the analysis of the dust characteristics, we aim to understand the root of their variability, and get a glimpse of the inner disc behaviour.} 
   {We employed a novel approach to measure the properties of the dips, by combining class III templates with Gaia multi-epoch photometry to construct the intrinsic photosphere of the objects. We measured several dip properties including the depth of the dips, near-infrared (NIR) excesses, and their optical depths as a function of wavelength.}
   {We record 20 significant dips that range in their dip properties and show no relation to one another. In almost all cases, the low optical depths and small NIR excesses are observed. Comparison of their optical depths with grain opacity models show that the dips can be explained by the presence of dust substructures containing processed grains obscuring their photospheres and/or their discs. These grain distributions can have maximum sizes as large as 20$\mu m$ and in many cases have almost grey-like extinction, while some require a strong scattering component.}
   {The findings highlight the extent of the irregularity of dippers, but also link it to the dust dynamics in the inner regions of the circumstellar discs. The dust substructures causing the variability require processed dust grains to be lifted above the disc into the line of sight. Possible lifting mechanisms including disc winds, unstable accretion columns, and disc warps are discussed.}


\titlerunning{The dance of dust}
\maketitle

\section{Introduction}\label{sec: intro}

Protoplanetary discs, known to be the nurseries of planets, play a crucial role in determining the formation conditions and early evolution of planetary systems. The creation of close-in planets, similar to our own, is particularly influenced by the innermost regions of the disc ($<$1 AU) and so it is here that valuable insights can be gained into the formation process \citep{Lissauer2011}. It is also in this region that material is accreted onto the central star \citep[e.g.][]{hartmann2016} and ejected through disc winds and outflows \citep{Frank2014}, processes that play an important role in the evolution of the star and its circumstellar disc. Observational evidence has routinely shown that the outer regions of the disc are highly structured \citep[e.g.][]{Long2018, Andrews2020, Benisty2023}; however, the challenging nature of studying the inner disc has still left the question of whether structures are present in the inner 0.5 au. These scales can only be marginally resolved with available observations \citep{Andrews2019}; however, what is known through photometric and spectroscopic surveys is that the region is highly complex and variable. 

Time variability of the continuum emission, as well as of the accretion and wind-tracing lines, can offer a unique insight into the inner disc properties \citep{Sicilia-Aguilar2020, CampbellWhite2021, Armeni2023, Sicilia-Aguilar2023}. For example, the variable light curve of AA Tau was interpreted to result from the occultation of the star by an inner disc warp induced by a misaligned magnetic field \citep{Bouvier2003}. Large photometric campaigns further revealed a class of young stars -- $\sim$30\% of the young stellar object (YSO) population -- the so-called dippers, with multiple brightness dips in their light curves \citep{Alencar2010a, Cody2014, Cody2018, Roggero2021, CampbellWhite2021}. Some objects have very regular, small variations, likely due to spots on the stellar surface.

Other objects  show  very irregular and  deep  dips (accounting for up to 20-50\% of the photospheric flux), likely due  to variable obscuration by circumstellar dust \citep{McGinnis2015}. Such events, whose origin is a matter of debate, are more likely related to transient changes  of the disc itself, or of the amount of dust lifted by winds \citep{Bans2012,Rodenkirch2022}, or entrained in the accretion columns.
Since in several cases these irregular dips occur in discs that have low to moderate inclination at large scales ($\leq$10-100 au) as probed by the Atacama Large millimetre/submillimetre array (ALMA), they may trace the presence of highly misaligned inner discs \citep{Ansdell2020,Scaringi2016}. Further evidence can be found in the shadows left by the misaligned inner disc onto the outer disc as observed in scattered light \citep{Benisty2017,Benisty2018}. In the case of the dipper RXJ1604-2130, the shadows vary in location and amplitude \citep{Pinilla2018}, with a short timescale that suggests that the obscuring dust is located very close to the star \citep{Sicilia-Aguilar2020}.

Considering the range of possible variability mechanisms, dip amplitudes, and frequency of stellar spot evidence, irregular dippers are therefore ideal targets to investigate the complex phenomena occurring in the inner disc. This can be done by comparing properties at different epochs, as the star suffers different degrees of obscuration. Time series of flux-calibrated spectra that simultaneously cover a broad range of wavelengths can provide unique information on accretion and wind properties, as well as on the location and properties of the obscuring grains \citep{Schneider2018}.

This paper presents the results of a survey of 16 irregular dippers in the star-forming regions Upper Scorpius, observed with X-Shooter mounted on the ESO's Very Large Telescope (VLT), focussing on the analysis of the continuum variability. The paper is structured as follows. In Section \ref{sec: sample_obs}, we present the VLT/X-Shooter observations and the sample properties. In Sect.\ref{sec: photosphere} we describe the method of determining the stellar photosphere used to compare the spectra of the objects in a dip. Section \ref{sec: dust} shows the first results of the survey along with the measured dust properties of the dips. Section \ref{sec: nir} investigates the near-infrared (NIR) emission of the objects, while \ref{sec: accretion} looks at the line veiling of the dips to search for any connection to variable accretion. In Section \ref{sec: discussion} we discuss the findings and their significance for such objects, which is followed by our conclusions in Section \ref{sec: conclusions}.

\section {Sample and observations}\label{sec: sample_obs}

\subsection{Sample properties}
The sample comprises of 16 objects from the Upper Scorpius star forming region. They are classified as irregular dippers by \cite{Ansdell2016, Ansdell2018, Ansdell2020}  and \cite{Hedges2018}. We selected objects with frequent and deep dips (median dip depth of 20\%), characteristics of irregular dips. The overall sample represents $\sim$ 35\% of the known dipper population in the region.   \\
The selected objects are all low mass, late spectral type (SpT) stars, mostly K and  M, with just one G-type star. Their stellar parameters are shown in Table~\ref{tab: master_table}. This table reports also the disc inclination as measured by ALMA \citep{Carpenter2024, Ansdell2020, Barenfeld2017}. In general, the selected objects have relatively high inclination angles ($i$) varying from 50\degr\ to 60\degr. Only two objects have measurements of $i \le \ang{20}$ (namely J155836 and J160421). These objects have been reported to have misaligned discs \citep{PerezLaura2018, Benisty2018, Nealon2020, Mayama2018}. See \Cref{sec: AppC_cases} for more details.

\begin{table*}
\begin{center}
\caption{Stellar parameters.} 
\label{tab: master_table}
\begin{tabular}{@{} c c c c c c c  c@{}}
\hline
\hline
 2MASS ID    & EPIC ID &       SpT & Dist      & $A_{V}$   & $M_*$  & Log $L_*$  & $i$               \\
             &         &        & (pc)      &           & ($M_\odot$) & ($L_\odot$) & (deg)  \\ 
\hline
\\
 J15554883-2512240 & 203710077 & G8$^{j}$          & 143$^{g}$                 & 0.7$^{c}$  & 1.35$^{c}$        & 0.37$^{c}$   &  ...                                                \\
 J15575444-2450424$^{*}$ & 203810851 & K4$^{b}$           & 134$^{g}$     & ...    &  ...          &   ...         & ...                                              \\
 J15583692-2257153$^{*}$ & 204281213 & K0$^{a}$          & 167 $^{g}$       & 0.0$^{a}$    & 1.63$^{a}$        & 0.4$^{a}$        & $19^{+4.0}_{-4.0}$ $^{h}$                                 \\
 J16020517-2331070$^{*}$ & 204137184 & M4.5$^{b}$         & 147$^{f}$        & 0.41$^{e}$ & ...           &  ...           &   ...                                         \\
 J16020757-2257467$^{*}$ & 204278916 & M2$^{a}$          & 140 $^{g}$      & 0.4$^{a}$  & 0.44$^{a}$        & -1.10$^{a}$        & $56.2_{-12.1}^{+11.3}$ $^{k}$                       \\
 J16041893-2430392$^{*}$ & 203895983 & M2$^{a}$          & 145$^{a}$            & 0.3$^{a}$  & 0.37$^{a}$        & -0.35$^{a}$        & $63.6^{+4.0}_{-4.7}$ $^{k}$                              \\
 J16042097-2130415 & 204637622 & M3.5 $^{b}$ & 149 $^{f}$ &  ... & ... & ... & ... \\
 J16042165-2130284$^{*}$ & 204638512 & K3$^{a}$          & 145 $^{g}$      & 1.4$^{a}$  & 1.24$^{a}$        & -0.05$^{a}$         & $7.8_{-0.1}^{+0.1}$ $^{h}$                                   \\
 J16064794-1841437$^{*}$ & 205238942 & M0$^{b}$           & 153$^{g}$          & 0.8$^{c}$  & 0.56$^{c}$        & -0.17$^{c}$  & $55.5^{+0.1}_{-0.1}$ $^{h}$                                 \\
 J16072747-2059442$^{*}$ & 204757338 & M4.75$^{b}$       & 179$^{c}$          & 0.5$^{c}$  & 0.13$^{c}$        & -0.89$^{c}$  & $83.9_{-35.0}^{+5.3}$ $^{k}$                                   \\
 J16090075-1908526$^{*}$ & 205151387 & M0$^{a}$          & 136 $^{g}$      & 1.0$^{a}$    & 0.6$^{a}$         & -0.49$^{a}$        & $54.3^{+5.4}_{-5.5}$ $^{k}$                 \\           
 J16115091-2012098$^{*}$ & 204932990 & M3.5$^{b}$        & 135$^{g}$          & 0.3$^{c}$  & 0.20$^{c}$        & -1.15$^{c}$  & $86_{-42}^{+4}$ $^{d}$                                    \\
 J16141107-2305362 & 204245509 & K4$^{a}$         & 145$^{a}$             & 0.3$^{a}$  & 1.25$^{a}$        & 0.02$^{a}$        & $59_{-17.3}^{+29.7}$ $^{k}$                                   \\
 J16154416-1921171 & 205110000 & K5$^{b}$           & 126$^{c}$          & 2.0$^{c}$  & 0.66$^{c}$        & -0.38$^{c}$  & $50.7_{-23.8}^{+18.6}$ $^{k}$                                   \\
 J16221852-2321480$^{*}$ & 204176565 &  ...          & 135$^{g}$  &  ...   & ...           &  ...          & $47.2^{+1.0}_{-1.0}$ $^{h}$                                  \\
 J16264741-2314521 & 204206295 &   ...               & 133 $^{g}$        & ...    &  ...          &  ...          & ...                                             \\
 
\hline
\end{tabular}
\tablebib{(a)~\citet{Manara2020};
(b) \citet{Luhman2020}; (c) \citet{Fang2023}; (d) \citet{Barenfeld2017};
(e) \citet{Ansdell2016}; (f) \citet{Jonsson2020}; (g) \citet{GaiaDR3};
(h) \citet{Ansdell2020} (j) \citet{Torres2006} (k) \citet{Carpenter2024}
}
\end{center}

\end{table*}

\subsection{VLT/X-Shooter observations and data reduction}
The data were taken with the X-Shooter spectrograph \citep{vernet2011}. This work combines spectra from a collection of observing programmes, spanning from 2016 to 2021 (see Table\,\ref{tab: logobs}). The first programmes (Id. 097.C-0378, 0101.C-0866 and 0103.C-0887; PI Manara, observations in 2016, 2018 and 2019, respectively) account for several epochs of four objects (J155836, J160900, J160421, and J160207). The most recent programme (Id. 0105.C-0513(A); PI Manara) obtained in 2021 contains two to four spectra on additional 12 objects completing the dataset. The time interval between observations vary from more than a year to just one day. Therefore for all stars, we obtained 2-4 spectra, with no specific strategy in terms of time separation. This method of stochastic observations maximises the probability of catching an object in a dipping state while also allowing the combination of archival and new data. The same observing strategy was used in all programmes. More details on the  observations and data reduction for the data from 2016 and 2018 can be found in \cite{Manara2020}, where a detailed study of the accretion properties was completed. Here the analysis was done using only the first epoch for each target. 

X-Shooter can obtain simultaneous observations in its three arms, (UVB, VIS, and NIR) covering the wavelength range 300\,nm--2500\,nm, allowing for a broad wavelength coverage of the dipper variability. The spectra were obtained with narrow (1\farcs0, -0\farcs4, -0\farcs4) slit widths, leading to spectral resolutions of $\gtrsim$ 5,000 for the UVB, 18,000 for the VIS, and 10,000 for the NIR arms. Additional spectra with wide (5\arcsec) slit widths were taken immediately after for flux calibration purposes. This was to ensure no flux losses due to varying seeing conditions. The slits were aligned at the parallactic angle to minimise atmospheric dispersion. The combination of the data leads to a database containing a total of 16 sources with two to four epochs per source (a total of 46 spectra).

Standard data reduction was completed using the X-Shooter pipeline v3.6.1 \citep{modigliani2010} with EsoReflex v2.11.5. The image reduction and analysis facility (IRAF) was used to extract the spectra in each arm (see \citealt{Manara2020} for more details). After standard flux-calibration procedures, the spectra taken with the narrow slit widths were scaled to the flux calibrated spectra obtained with the wide slit set-up. The wavelength calibration from the pipeline was refined by using the \LiI\ (6707.856\,\AA) absorption line characteristic of the photosphere of late type low mass stars \citep{Campbell-White2023}. Any remaining noise in the continuum was simply cut through high threshold (17$\sigma$) sigma clipping. No telluric correction was performed, regions effected by atmospheric interference were cut from the spectra. The final sample of reduced flux calibrated spectral energy distributions are shown in Appendix \ref{sec: AppA_data}.

\section{The stellar photosphere}\label{sec: photosphere}

The reduced and flux-calibrated spectra (Sect.\,\ref{sec: AppA_data}) show the characteristic spectral energy distributions (SED) of Class II objects with little to moderate infrared excess in this wavelength coverage. The majority of objects have multi-epoch SEDs showing significant differences in their continuum levels. This is the dipping behaviour seen from a spectroscopic point of view. In some cases the spectra also show different amounts of NIR excess between epochs. 

In order to characterise the dipping events and probe their nature, the 'intrinsic' spectrum of the object (outside dimming events) needs to be identified and characterised. This is done by defining the inherent photospheric flux of the source, along with its spectral type. In the following we describe how the source spectral type and intrinsic continuum flux have been derived.

\subsection {Spectral type}

The spectral type of the objects in the sample was derived by comparing the observed spectra to a collection of Class III spectra following \cite{Manara2013, Manara2017, Claes2024}. As described in this work, Class III objects are preferred over synthetic spectra as the latter do not account for the high chromospheric activity characteristic of low gravity objects such as pre-main sequence stars. The Class III sample was selected from \cite{Claes2024} \footnote{Templates can be accessed on github at https://github.com/RikClaes/FRAPPE}. As in our case, these spectra were observed with X-Shooter, and they were reduced and analysed in the same fashion. 
To find the best match between templates and observed spectra we considered the position and depth of the molecular bands in the visual at 705\,nm--706\,nm, 708\,nm--709\,nm, and 712\,nm--713\,nm in the case of M-type stars. For earlier spectral types, the determination of the SpT is based on a Mg-line index \citep{Herczeg&Hillenbrand2014, Claes2024}. The overall continuum shape and presence of a NIR excess was also taken into account when determining the spectral type. To determine the uncertainty in the SpT, the molecular bands and previously mentioned spectral index of the same object were derived at different epochs. On average, the depth of the molecular bands used in the spectral type determination of M-type stars vary by $\lesssim 5$\% in six out of the eight objects, and by $\lesssim 15$\% in the remaining two. For earlier spectral types, variations of $\lesssim $ 10\% are found. In all cases, this translates into an uncertainty of the SpT of less than two subclasses. 

The derived spectral types are reported in Table~\ref{tab: template_results}, together with the adopted Class III template. The differences with previous estimates are always within three subclasses. It should be noted that in contrast with our study, previous spectral type determinations of our objects may have been performed from single epoch observations, and therefore, local variations in the extinction -- causing the dipping events, not related to the interstellar medium (ISM) -- were not taken into account. This can account for the reported differences.

Using the method described above, suitable templates were found for all our sources with the exception of two objects: J162647 and J161544. Both objects show a very red SED, with J161544 being potentially a misclassified Class I object. Both objects are excluded from further analysis.

\subsection{Normalisation of the class III template and interstellar extinction}\label{sec: classIII_extinction}
Once the SpT is determined, the template spectra need to be calibrated to the intrinsic photospheric flux. Multi-epoch observations of dippers show that their fluxes decrease multiple times from a maximum flux level that is seldom overreached. Unfortunately, the sparse and sporadic time coverage of our data do not allow us to determine the maximum (or quiescent) flux level of our sources. However, for 13 out of 16 objects, Gaia offers time-sequenced photometric measurements \citep{Gaia2016, GaiaDR3}. We can make use of this multi-epoch Gaia photometry to determine the quiescent flux level of our sources. Furthermore as Gaia optical observations are free from excess emission due to the presence of, for example, a disc, one can assume the quiescent flux level as the intrinsic photospheric flux of our sources. This assumes that the observed variation in the flux level of our spectra at optical wavelengths is due to a variable extinction due to the sum of line-of-sight extinction, that should be constant in time, and of a local, unknown contribution, variable in magnitude and wavelength dependence, which causes the dips.

To obtain the photospheric flux level of our sources and the ISM reddening, synthetic G, BP and RP magnitudes were estimated from our Class III templates. The templates were reddened by the ISM extinction using \cite{Cardelli89} with $R_V=3.1$. The ISM extinction was varied until there was a good match between the BP and RP magnitudes normalised to the Gaia G band magnitude at quiescence, and the reddened synthetic magnitudes of our Class III template.    
It should be noted that normalising to the broad band, Gaia G filter, also corrects for the distance to the objects. \Cref{fig: gaia_magdiff} in Appendix \ref{sec: AppB_Gaia} plots the difference between the observed Gaia magnitudes and the synthetic magnitudes of the adopted photospheric flux. The derived ISM extinction values are included in \Cref{tab: template_results}. The derived values agree with those reported in the literature within typically $\Delta A_V = 0.5$ mag. The only exception is J160421 where a difference of 0.8\,mag is found.

\begin{table}
\caption{Spectral type and extinction derived in this work.}
\scriptsize
\centering
\begin{tabular}{@{} c c c c c c @{}}
\hline
\hline

2MASS ID  & Class III        & SpT & $\Delta$ SpT &$A_V$  & $\Delta A_V$\\

\hline \\

J155548 & RXJ1508.6-4423          & G8  &  0      & 0.3  &  0.4    \\
J155754 & MTLup                   & K5  & 1         & 0.2  & ...      \\
J155836 & CD-31 12522            & K0   & 0        & 0.0    & 0.0      \\
J160205 & Sz94                    & M4  & 0.5         & 0.7  & 0.3      \\
J160207 & RXJ1121.3-3447 app1    & M1   & 1         & 0.6    & 0.2\\
J160418 & RXJ1121.3-3447          & M1 & 1          & 0.3     & 0.0   \\
J160420 & RX J160421-21307        & M3  & 0.5         & 0.3  &...      \\
J160421 & RXJ1547.7-4018          & K3  & 0         & 0.6    & 0.8    \\
J160647 & 2MASSJ15552621-3338232  & K5  &3         & 0.5     & 0.3   \\
J160727 & Par Lup3 2            & M5    &0.25       & 0.1    & 0.4    \\
J160900 & THA15-36A               & M0  &0         & 0.5     & 0.5   \\
J161150 & 2MASS J16115091-2012098 & M3  & 0.5        & 0.8   & 0.5     \\
J161411 & RXJ0438.6+1546          & K2  &2           & ...   & ...       \\
J161544 &  ...                   & ...  & ...           & ... & ...          \\
J162218 & RXJ1547.7-4018          & K3  & ...         & 1.9   &...     \\
J162647 &  ...                       & ... &...         & ..  & ...     \\

\hline
\end{tabular}

\tablefoot{$\Delta$ SpT, $\Delta A_V$ are the differences between those determined in this study and those reported in literature in \Cref{tab: master_table}. $\delta$ SpT values are expressed in terms of subclasses.}
\label{tab: template_results}

\end{table}

For two of the three objects  with no multi-epoch Gaia photometry (J162647 and J161544), the photospheric flux calibration was done in a similar fashion but using the available photometric measurements from the literature. These were compared with the available single epoch G-band photometry to determine the brightness state of the source. It should be noted that for these two sources, the resulting photosphere has a larger uncertainty than the rest of the sample, as the value of $A_V$ cannot be accurately constrained. Finally no representative Gaia photometry was reported for J161411, and therefore the photospheric level of the source could not be determined. This source was excluded from further analysis.

The reconstructed, reddened intrinsic continuum photospheric emission of each star is shown in Fig.\,\ref{fig: master_sed} in Appendix\,{\ref{sec: AppA_data}} alongside the observed spectra. Following the discussion above, the total sample included in this work contains 13 objects.

\section{Dust properties as traced by the dips}\label{sec: dust}

\begin{figure}
    \centering
    \includegraphics[width=\linewidth]{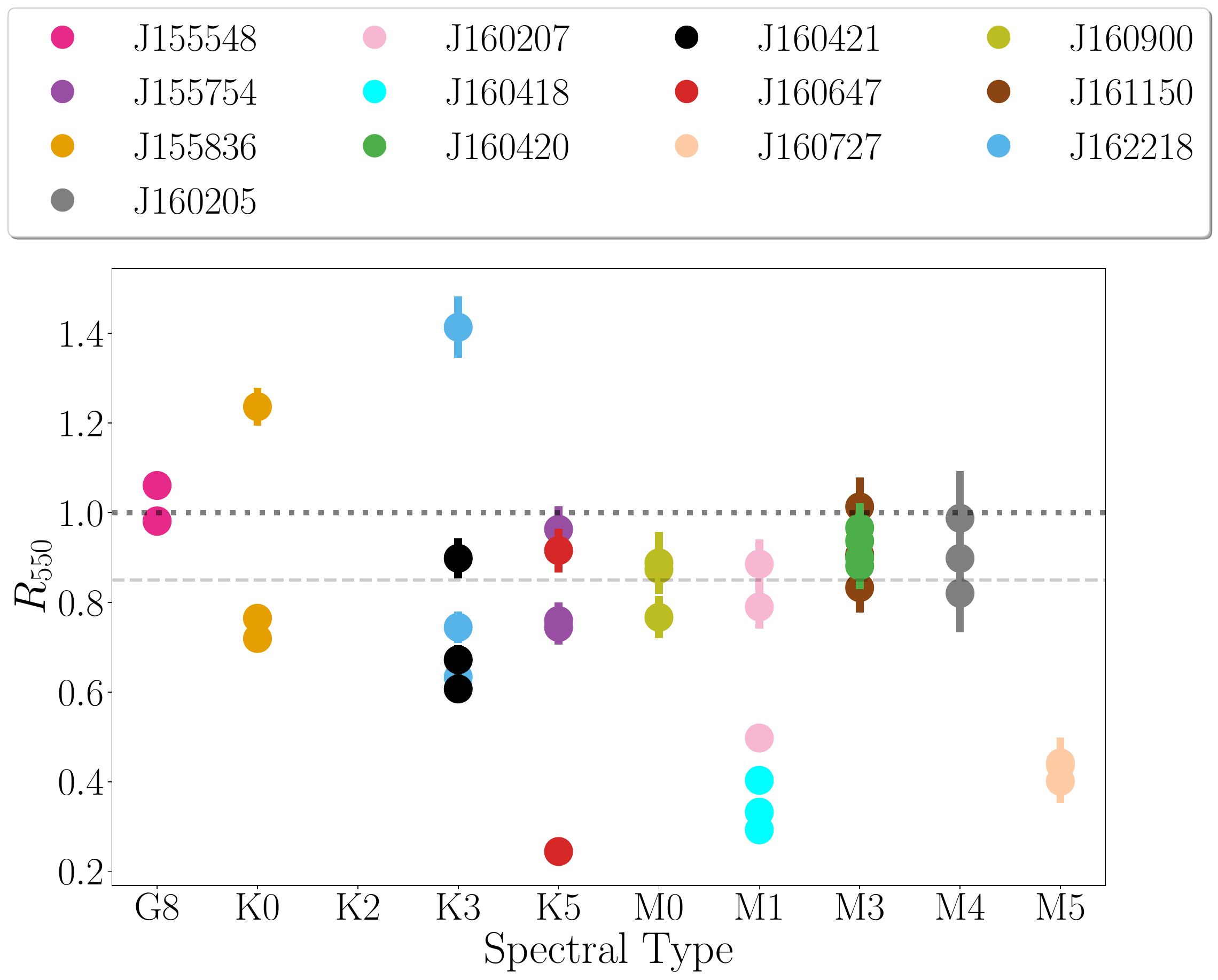} 
    \caption{Ratios of the observed to photospheric flux at 550 nm ($R_{550}$) for each star (see colour legend) are shown as a function of the spectral type. When not visible, the errors are smaller than the size of the dots. Dotted and dashed grey lines represent \Rvis thresholds of 1.0, 0.85 respectively. No relation between spectral type and \Rvis is observed.}
    \label{fig: dip_stats}
\end{figure}

In order to probe the continuum variations from epoch to epoch, the ratio of the ISM corrected observed spectrum to the photospheric one, $R_\lambda$, was estimated (see Fig.\,\ref{fig: gaia_R_masterplot}). The ratios were computed from the observed spectra smoothed with equally spaced bins in wavelength (95\,\AA), and sigma-clipped to remove any effects of emission lines. To better visualise the results, Fig.\,\ref{fig: dip_stats} shows the value of this ratio at $\lambda=550$ nm (\Rvis) computed for each multi-epoch observation. As shown in the figure, obscuring events are variable from source to source and epoch to epoch for the same source, with values generally ranging from \Rvis $\sim$1 (i.e. no dip) to \Rvis$\sim$0.5. Only 7 events (belonging to 4 different objects) show  \Rvis\ values $\le 0.5$. In some cases, dimming events are clustered around a specific value (e.g. J160418). Fig. \ref{fig: dip_stats} also highlights that some spectral types (e.g. K3, M1) have a wide spread of \Rvis values indicating no relation between spectral type and the dip ratios. On a separate note, there is no trend seen when comparing the time separation between observations and values of \Rvis. This confirms that our observing strategy samples randomly the dipping history of the objects, and, also, that these are indeed irregular dippers.

The most likely scenario for the presence of the observed dimming events is the presence of local dust, increasing the extinction towards the source, and/or the presence of cold spots on the stellar surface. Both of these scenarios will produce a decrease in the observed flux. Cold spots are generally short-lived and appear regularly in time, at least for a period, in phase with the stellar rotation. In order to relate the observed dimming events to the presence of cold spots, we analysed K2 curves \citep{Howell2014, Huber2016} of several of our objects in order to look for regular variability and assess the degree of flux variations. Based on our analysis, we find that regular dips, likely associated with cold spots, are unlikely to reduce the photospheric flux by more than $\sim$ 15\%. Therefore, dimming events above this limit are most likely due to variable local dust obscuration. In this section, we will focus on probing the properties of the dust causing the dimming events. Therefore, we will limit our study to events with \Rvis < 0.85, avoiding significant contamination from cold spot contributions. Taking this limit into account, our analysis will focus on 20 dimming events from a total of 11 sources (marked with a * in \Cref{tab: master_table}). 
 
\subsection {Wavelength dependence of the optical depth}\label{sec: beta}

The wavelength dependence of $R_{\lambda}$ provides very important clues on the properties of the local layer of dust causing the dimming events. From $R_\lambda$, it is possible to estimate the optical depth of the obscuring layer of dust as 

\begin{equation}
    \tau_\lambda = -\ln{R_\lambda}.
\end{equation}

A rough estimate of the optical depth dependence on the wavelength can be estimated by computing $\beta$, which represents the ratio of the optical depths at two reference wavelengths $\lambda_1$  and $\lambda_2$:

\begin{equation}
     \label{eq: beta}
     \beta= \frac{\log(\tau_2/\tau_1)}{\log(\lambda_2/\lambda_1)} .
\end{equation}
We choose to focus our measurement in the visual range of the spectrum ($\lambda_1 \approx 400\,$nm and $\lambda_2 \approx 900$\,nm). In doing so we avoid contamination from the lower signal-to-noise ratio (S/N) in the UVB, and the hot dust emission that dominates in the NIR. \\

\begin{figure}
    \centering
    \includegraphics[width=\linewidth]{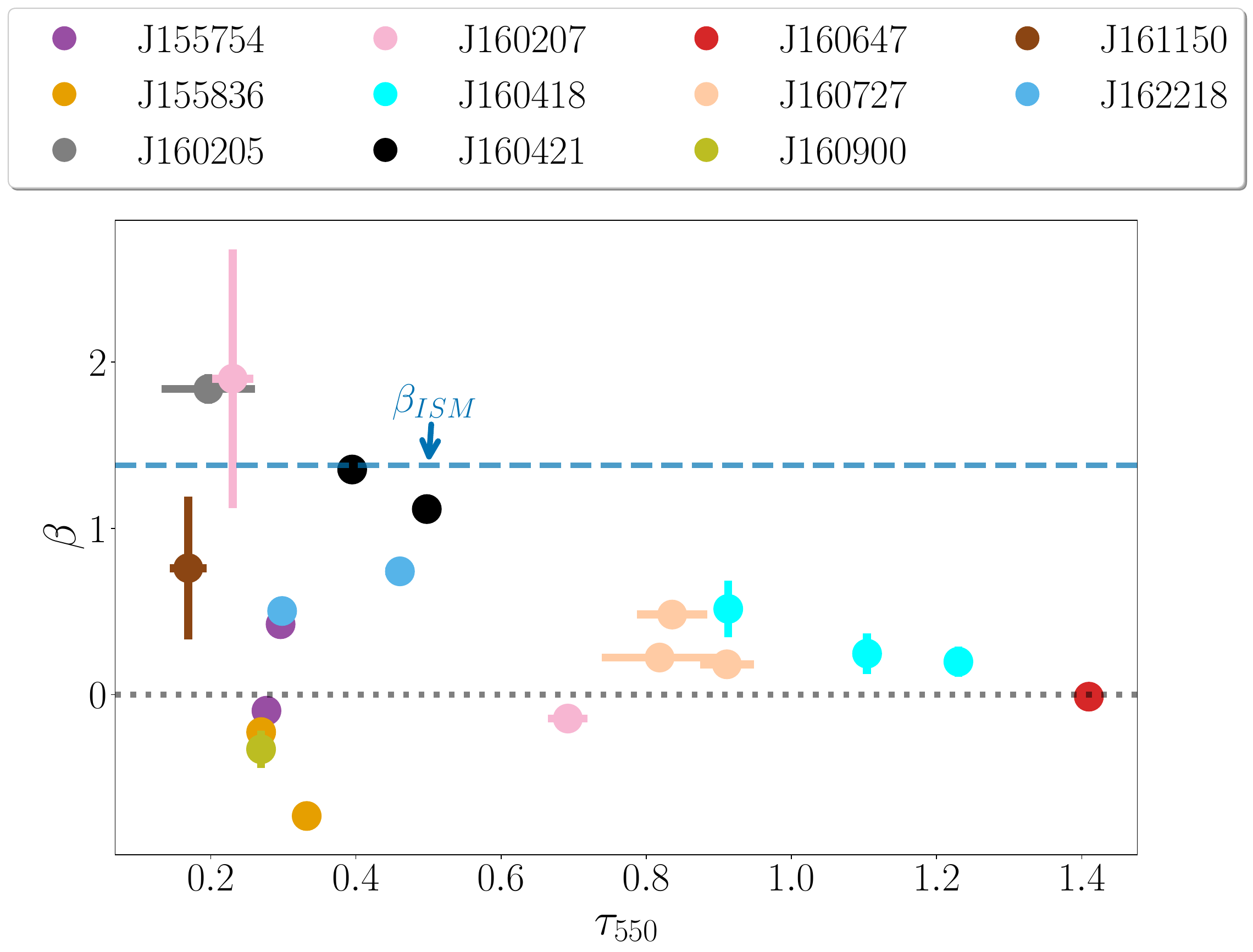}
    \caption{$\beta$ (eq. \ref{eq: beta}) values as a function of \tauvis ~for objects with \Rvis<0.85. Multiple dips of the same object are colour coded according their shortened 2MASS ID (see \cref{tab: master_table}). Dashed blue line represents the value expected for ISM reddening under typical conditions, assuming the Cardelli extinction law \citep{Cardelli89} with $R_V$=3.1. }
    
    \label{fig: beta_plot}
\end{figure}

The values of $\tau_1 , \tau_2$ were computed from the weighted mean of the three nearest values at $\lambda_1, \lambda_2$ from the binned $\tau_\lambda$ values, as can be seen in Fig \ref{fig: master_tau}. The uncertainties on the values of $\tau_1 , \tau_2$ are derived from the noise of the spectra in each bin of $R_\lambda$. These are carried through, using the standard method of error propagation, in computing $\tau_\lambda$, the weighted means, and ultimately the value of $\beta$. Note that $\tau_\lambda$ and $\beta$ refer to the local dust only as ISM extinction correction has already been applied as explained in Section \ref{sec: classIII_extinction} and reported in \ref{tab: template_results}. Figure\,\ref{fig: master_tau} shows $\tau_\lambda$ for the 20 dimming events associated with the presence of dust. Note that $\tau_\lambda$ and $\beta$ refer to the local dust only as ISM extinction correction has already been applied as explained in Section \ref{sec: classIII_extinction} and reported in \cref{tab: template_results}. figure~\ref{fig: beta_plot} shows the values of $\beta$ versus the optical depth at 550 nm, $\tau_{550}$. As a reference, the $\beta$ value corresponding to the ISM extinction taken from \cite{Cardelli89} assuming $R_V$ = 3.1 is also shown in the plot.

In practice, fig.~\ref{fig: beta_plot} shows a crude measure of the dependence of $\tau$ on $\lambda$ ($\tau \propto \lambda^{-\beta}$), and thus provides information about how the dust extinction (i.e. effective scattering and absorption opacities; $k_\lambda$) varies between $\lambda_1, \lambda_2$ for each dip. As shown in the figure, $\beta$ values are always lower, and in many cases significantly lower, than that expected from ISM reddening, except for two dimming events (for objects J160207 and J160205). It should be noted that five dips show negative $\beta$ values implying that the dust extinction increases with increasing wavelength. This is the opposite behaviour expected from ISM extinction, and it shows a more extreme behaviour than that expected for large grain size distributions (i.e. grey extinction, $\beta \sim 0$; \citealt{Draine2003}).

\subsection {Dust grain models}\label{sec: grainprop}

To further interpret the dependence of $\tau$ on wavelength and gain some insight into the nature of the dust causing the dimming events, dust opacity models were computed using the Optool code \citep{optoolCode}. The code performs opacity calculations for various dust grain properties and conditions. It should be noted that the analysis presented here is not intended as an exhaustive study of the dust properties of the disc, but as a tool to give rough estimates of the dust properties producing the dimming events and the relative dust property variations between epochs. 

\begin{figure}
    \centering
    \includegraphics[width=0.9\linewidth]{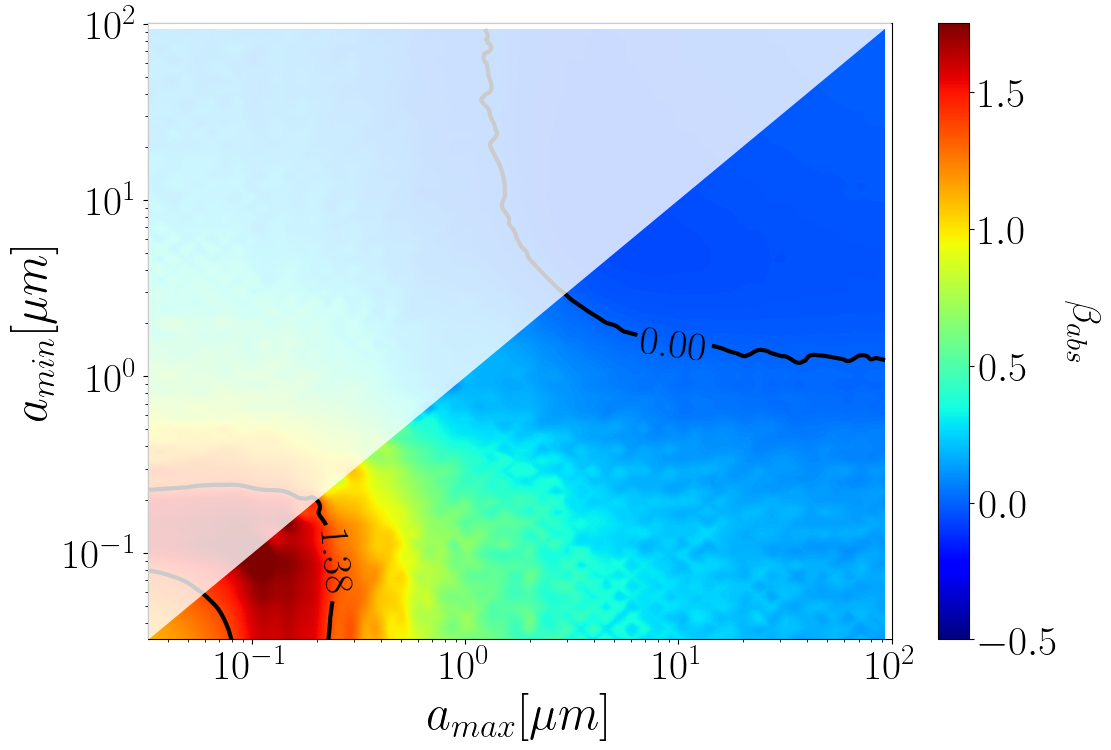} 
    \includegraphics[width=0.9\linewidth]{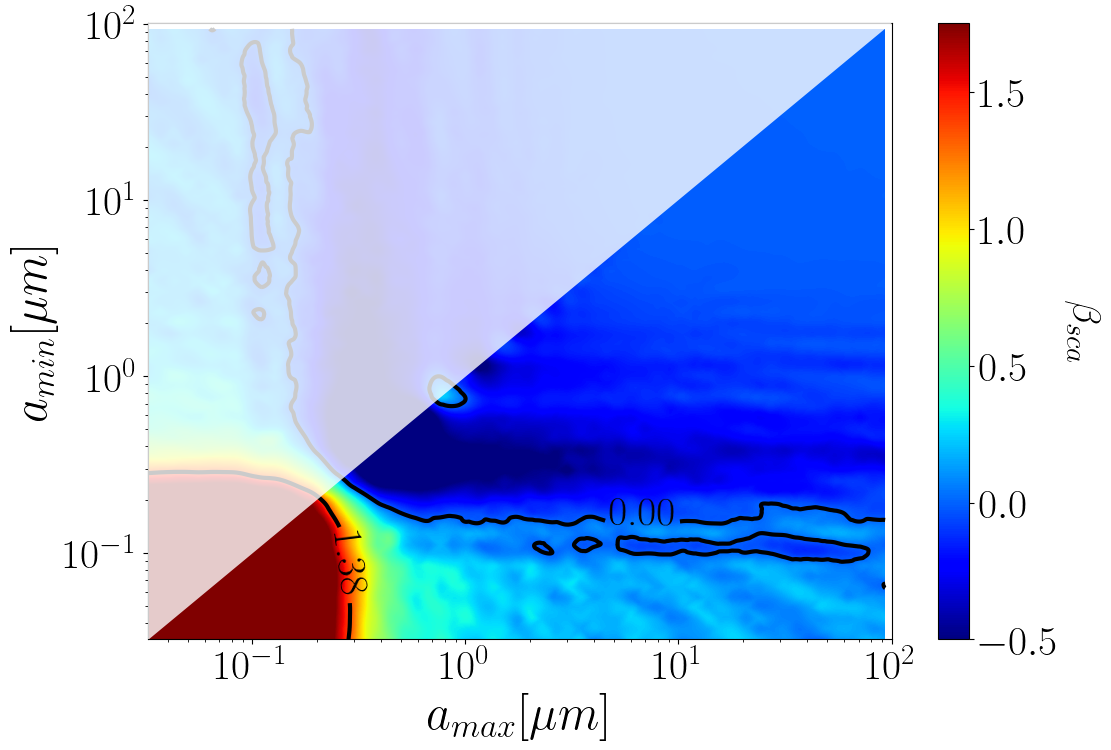}
    \includegraphics[width=0.9\linewidth]{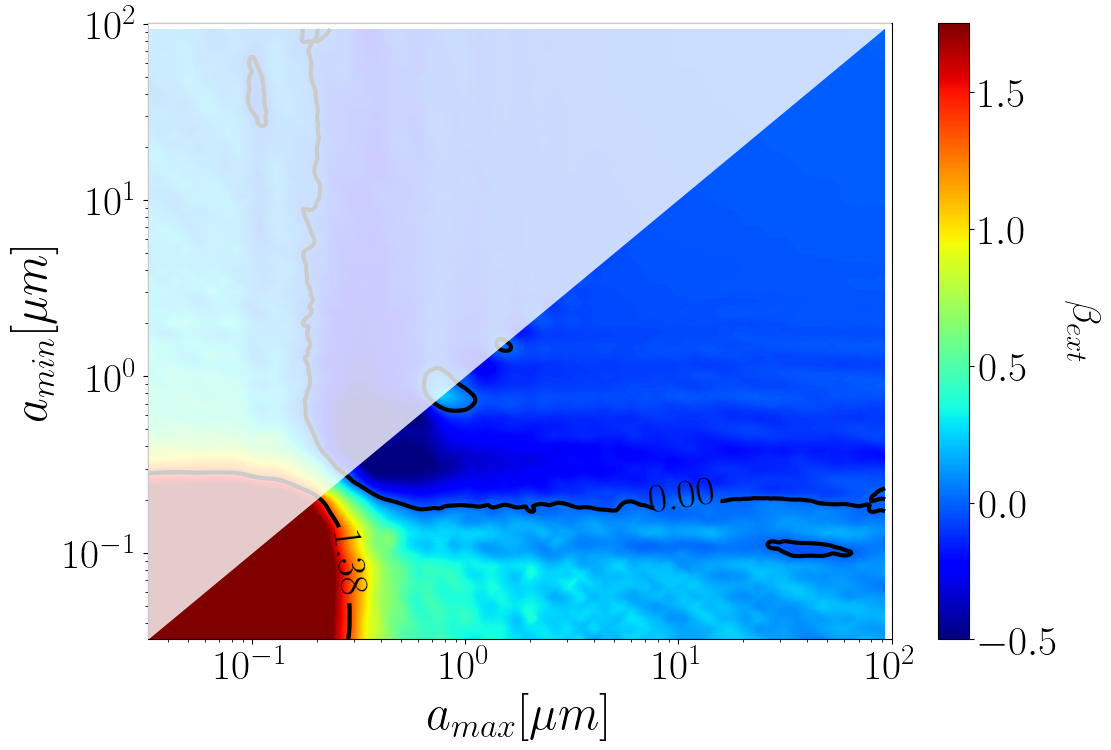}
    \includegraphics[width=0.9\linewidth]{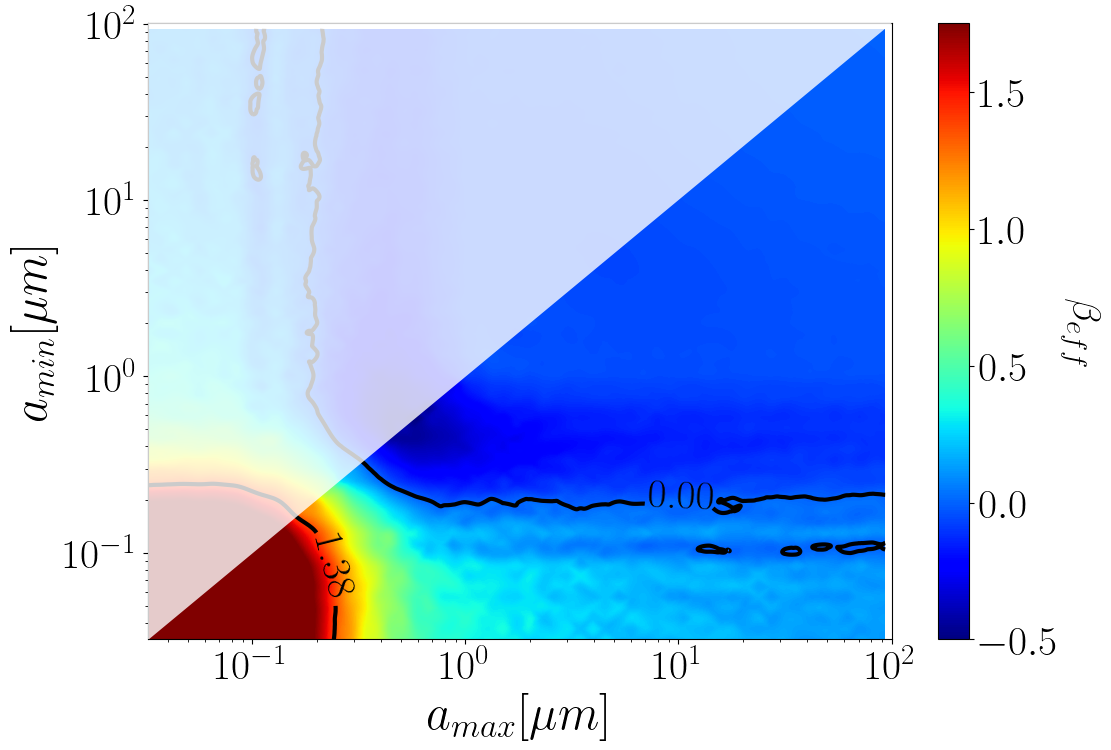}
    \caption{$\beta$ colour maps  from dust opacity models. Each figure represents a grid of dust size distributions varying  minimum and maximum grain sizes ($n(a)\propto a^{-q}$ with q=3.0). Coloured by the values of $\beta$ derived from the opacities for a) pure absorption ($\beta_{abs}$), b)scattering ($\beta_{sca}$, c) extinction ($\beta_{ext}$), and d) effective extinction ($\beta_{eff}$; see text). Contours are marked by black lines at $\beta$ =  0.0 and $\beta_{ISM} =1.38$). Pattern effects  are artefacts of the interpolation used. }
    \label{fig: dustmaps}
\end{figure}

We consider grains composed of composite silicate materials (astrosilicates; \cite{Draine2003}) with a size distribution $n(a) \propto a^{-q}$ between a minimum and a maximum size $a_{min}$ and $a_{max}$. The grains are assumed to be compact and spherical. We compute absorption ($k_{abs}$) and scattering ($k_{sca}$) opacities, and the phase function ($<g>$) as a function of wavelength. Note that we primarily consider single scattering when computing $k_{sca}$ and $<g>$. However, considering pure absorption, $k_{abs}$ only, can be interpreted as the extreme case of multiple scattering. For a first comparison to the observations, we characterise each curve by the corresponding value of the quantity $\beta$,  defined as in sect.\,\ref{sec: beta}. We vary $a_{min}$ and $a_{max}$ between 0.03\,\micron\ and 100\,\micron\ and fix  $q=3$. \Cref{fig: dustmaps} (first and second panel from top) shows the values of $\beta$ for absorption and scattering opacities separately.  In reality, the actual attenuation of the stellar radiation due to the intervening dust is a complex function not only of the dust properties but also of the geometry of the system \citep{Lakeland2022}. Two additional examples are provided in the lower panels of \cref{fig: dustmaps}, showing the values of $\beta$ for extinction ($k_{ext}=k_{abs}+k_{sca}$), and for the case where the contribution of scattering is weighted by the phase function ($\kappa_{eff}=\kappa_{abs}+\kappa_{sca}(1-<g>)$; \cite{Birnstiel2018}).

As shown in fig.\,\ref{fig: dustmaps}, distributions with both $a_{min}$ and $a_{max}\lesssim 0.3\,\mu m$ result in high values of $\beta$ ($ \gtrsim 1.4$). If absorption dominates, a wavelength-independent attenuation ($\beta\sim 0$), as detected in several spectra, can only be reproduced if the smallest grains (i.e. $a_{min}\gtrsim 1-2$\,\micron) are eliminated. The need to significantly increase $a_{min}$ remains if (single) scattering is included, even if less extreme. In the case of extinction, it is possible, for example, to have $\beta\sim 0$ with  $a_{min} \gtrsim 0.2$, as long as $a_{max} \gtrsim$ 1-2\,\micron. 
The very interesting cases of negative $\beta$, as measured in some spectra, can only be explained if scattering plays a significant role in the attenuation, and only if the grain size distribution is very narrow, limited to the range from few tenths to $\sim 1$\,\micron\ (dark blue areas in fig.\,\ref{fig: dustmaps}). Note that, because of the range of wavelengths used in the definition of $\beta$ (i.e. 400-900\,\micron), the results are in all cases not sensitive to values of $a_{min}\ll 0.3$\,\micron\ and $a_{max}\gg 10$\,\micron, respectively.

Although the trends discussed above are likely to be qualitatively correct, similar values of $\beta$ can be obtained by different sets of parameters even for a specific grain model. In addition, it is difficult to account for the effect that the uncertainties on $\beta$ have in the interpretation of fig.\,\ref{fig: dustmaps}. Moreover, it should be noted that $\beta$ provides a very crude description of the actual wavelength dependence of the observed optical depth, as it is based on two reference wavelengths only (see equation \ref{eq: beta}). With these considerations in mind, we examined each of the 20 $\tau (\lambda)$ curves seen in fig.\,\ref{fig: master_tau}. For each $\tau(\lambda)$, we derived the best fitting values of the grain size distribution parameters over the whole range between 400\,nm and 900\,nm, assuming, for simplicity, that the observed optical depth is due to extinction. As we are interested in the possibility that small grains (e.g., <1 \micron) do exist in the obscuring dust, we explore the full range of $a_{min}\ge 0.03$\,\micron, decreasing $q$ when required. This has the effect of making the grain size distribution more top heavy. The results are shown in fig.\,\ref{fig: master_tau} and summarised in table \ref{tab: grainsizes} . Note that we do a visual fit only as we consider that the unknown geometry and the uncertainties in the actual grain composition will make any statistical estimate of the parameters' errors meaningless.

\begin{table}
\begin{center}

\caption{Dust grain distribution parameters.}
\label{tab: grainsizes}
\begin{tabular}{c c c c c c}
\hline 
\hline
Name    & Epoch & q   & $\mathbf{a_{min}}$ & $\mathbf{a_{max}}$ & $\mathbf{\beta}$ \\
        &       &     &     (\micron)      &     (\micron)      &                   \\  
\hline
        &       &     &      &       &                \\
J155754 & 1     & 3.0 & 0.03 & 1.00  & $0.42 \pm 0.05$\\
 & 3     & 3.0 & 0.03 & $\ge$ 5.00  & $-0.10 \pm 0.02$\\
J155836 & 2     & 3.0 & 0.20 & 0.70  & $-0.73 \pm 0.06$\\
 & 3     & 3.0 & 0.20 & 0.45  & $-0.23 \pm 0.08$\\
J160205 & 1     & 3.0 & 0.03 & 0.25  & $1.84 \pm 0.09$\\
J160207 & 1     & 3.0 & 0.03 & $\ge$ 10.00 & $-0.14 \pm 0.06$\\
 & 3     & 3.0 & 0.03 & 0.40 & $1.90 \pm 0.78$\\
J160418 & 1     & 3.0 & 0.03 & 1.50  & $0.52 \pm 0.17$\\
 & 2     & 3.0 & 0.10 & 3.00  & $0.25 \pm 0.12$\\
 & 3     & 3.0 & 0.10 & 3.00  & $0.20 \pm 0.09$\\
J160421 & 1     & 3.0 & 0.03 & 0.35  & $1.12 \pm 0.06$\\
 & 3     & 3.0 & 0.03 & 0.30  & $1.35 \pm 0.04$\\
J160647 & 1     & 2.5 & 0.03 & $\ge$ 20.00 & $-0.01 \pm 0.00$\\
J160727 & 1     & 3.0 & 0.03 & 1.00  & $0.48 \pm 0.02$\\
 & 2     & 3.0 & 0.03 & 10.00  & $0.18 \pm 0.01$\\
 & 3     & 3.0 & 0.03 & 10.00  & $0.22 \pm 0.02$\\
J160900 & 2     & 3.0 & 0.03 & $\ge$10.00 & $-0.33 \pm 0.11$\\
J161150 & 3     & 3.0 & 0.03 & 1.00  & $0.76 \pm 0.43$\\
J162218 & 1     & 3.0 & 0.03 & 1.00  & $0.50 \pm 0.06$\\
 & 3     & 3.0 & 0.03 & 1.00  & $0.30 \pm 0.02$\\
\hline
\end{tabular}
\end{center}
\end{table}

\begin{figure*}
    \centering
    \sidecaption
        \begin{minipage}[c]{0.7\textwidth}
        \includegraphics[width=\textwidth]{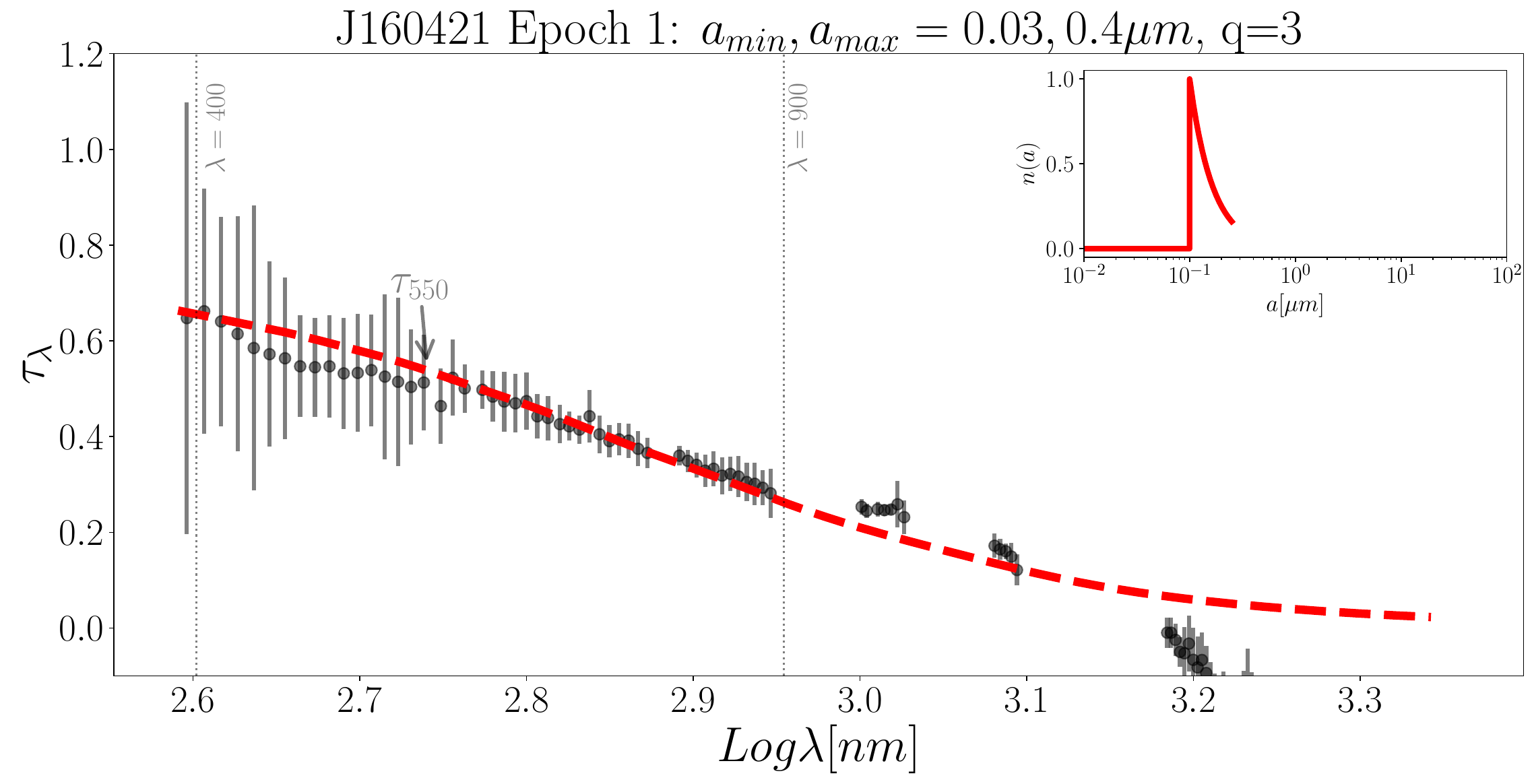}\\[1ex]
        \includegraphics[width=\textwidth]{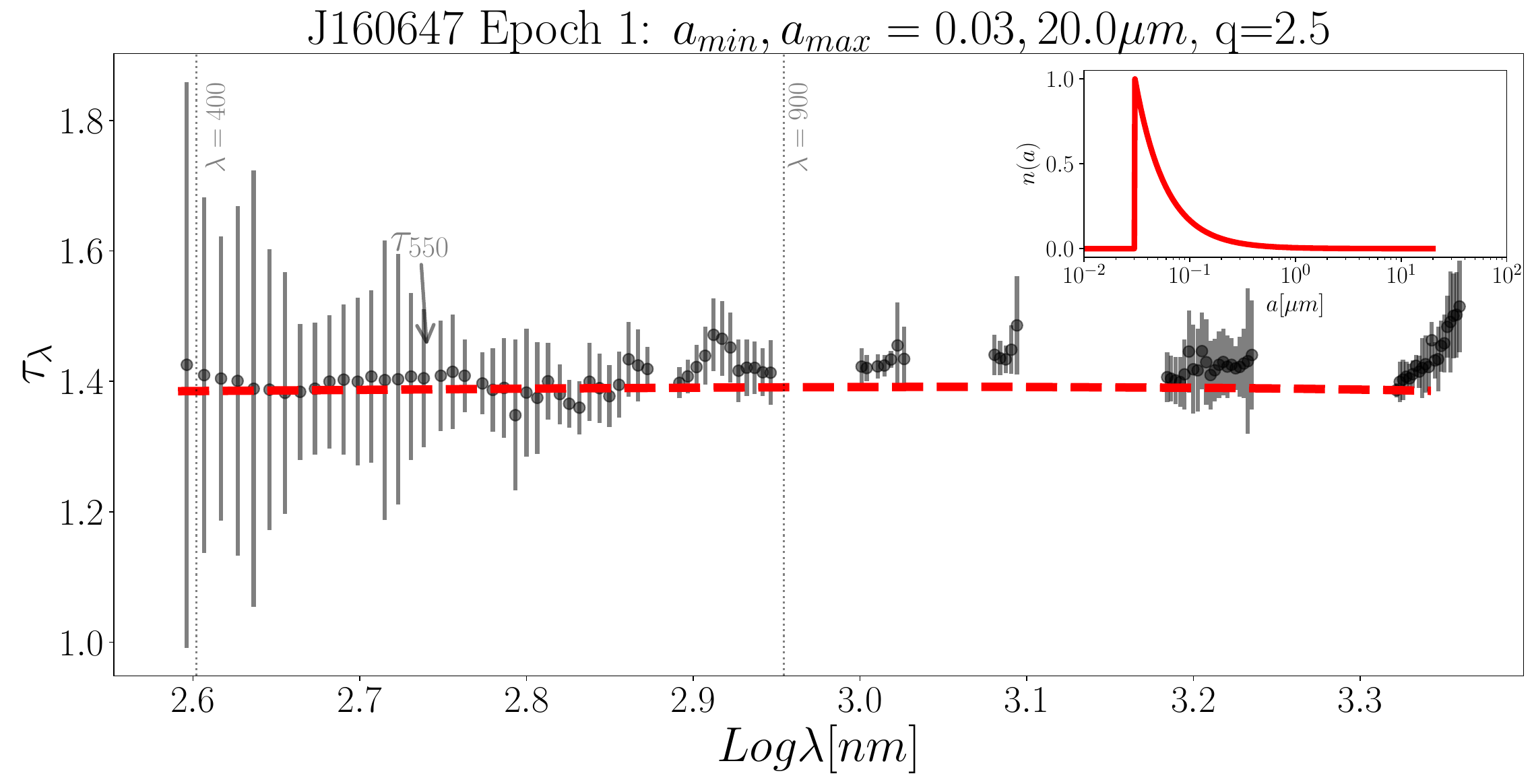}\\[1ex]
        \includegraphics[width=\textwidth]{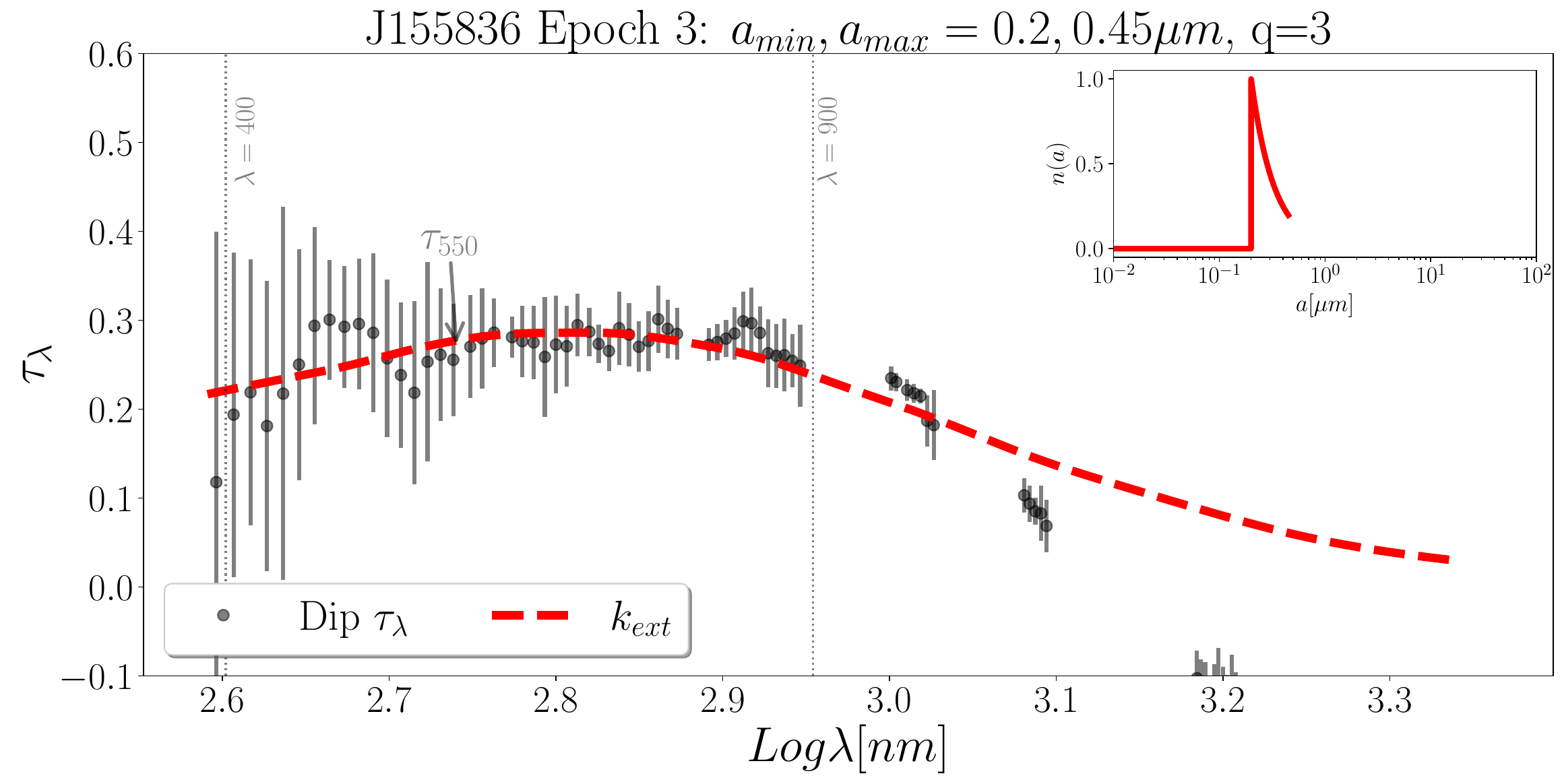}
        \end{minipage}
        \caption{Optical depth as function of wavelength for three representative dips. The observed values are shown by grey dots with error bars. The red dashed line shows the extinction opacity (absorption + scattering) for the grain size distribution that best fit the data in the wavelength interval 400-900 nm (limits shown by the vertical grey dotted lines), normalised to the observed value at 700 nm. The values of $a_{min}, a_{max}, q$ are given on top of each panel; the corresponding grain size distributions ($n(a)$) are displayed in the inset plots.}
    
        \label{fig: dust_and_taus}
\end{figure*}

Inspection of figs.\,\ref{fig: dustmaps}, \ref{fig: master_tau}, and table\,\ref{tab: grainsizes} suggest that the observed $\tau(\lambda)$ curves can be classified into three different groups, based on their wavelength dependence and matching grain size distributions. \Cref{fig: dust_and_taus} shows the three characteristic examples of this behaviour: decreasing $\tau(\lambda)$ with increasing wavelength (top panel); very flat $\tau(\lambda)$ (middle panel); curves not well described by a power-law in the  Log$\tau$-Log$\lambda$ plane (bottom panel). The curves shown in the top panel (i.e. decreasing $\tau(\lambda)$ with increasing wavelength) can be reproduced by extinction curves with $q=3$, $a_{min} \sim 0.03 \mu m$, and $a_{max}$ ranging from tenths to micron size particles. In some cases (e.g. J160205), $a_{max}$ is also small, and the grains have a size distribution similar to the ISM one.
The curves with very flat $\tau_\lambda$ ($\beta \sim 0.0)$ (fig.\,\ref{fig: dust_and_taus}; middle panel) are consistent with very small values of $a_{min} \sim 0.03$\,\micron, provided that $a_{max}$ is very large, of the order of a few tenths of micron or more. Finally, the example of curves not well described by a power-law in the Log$\tau$-Log$\lambda$ plane (fig.\,\ref{fig: dust_and_taus}; bottom panel) often results in $\beta<0$, and they require a narrow size distribution of grain sizes, in the tenth of micron range. Note that these kinds of curves require significant contributions of scattering to the total opacity. In this case, the albedo ($\omega = k_{sca} / k_{ext}$) at $\lambda =550$\,nm increases from 0.60 to 0.86 from the example shown in the middle to the bottom panel. 

In general, very few dips show evidence of ISM-type grains. Typically the dips sample in this study require flat $\tau(\lambda)$ curves, typical of grain size distributions with $a_{max}$ much larger than in the ISM, reaching in several cases, sizes of >10\,\micron. The minimum size is often consistent with ISM values, but in all cases the grain size distribution is shifted to much larger sizes, providing strong evidence that the grains in the occulting region come from a population that has undertaken significant growth.

\section{Near-infrared emission} \label{sec: nir}

\begin{figure}
    \centering
    \includegraphics[width=\linewidth]{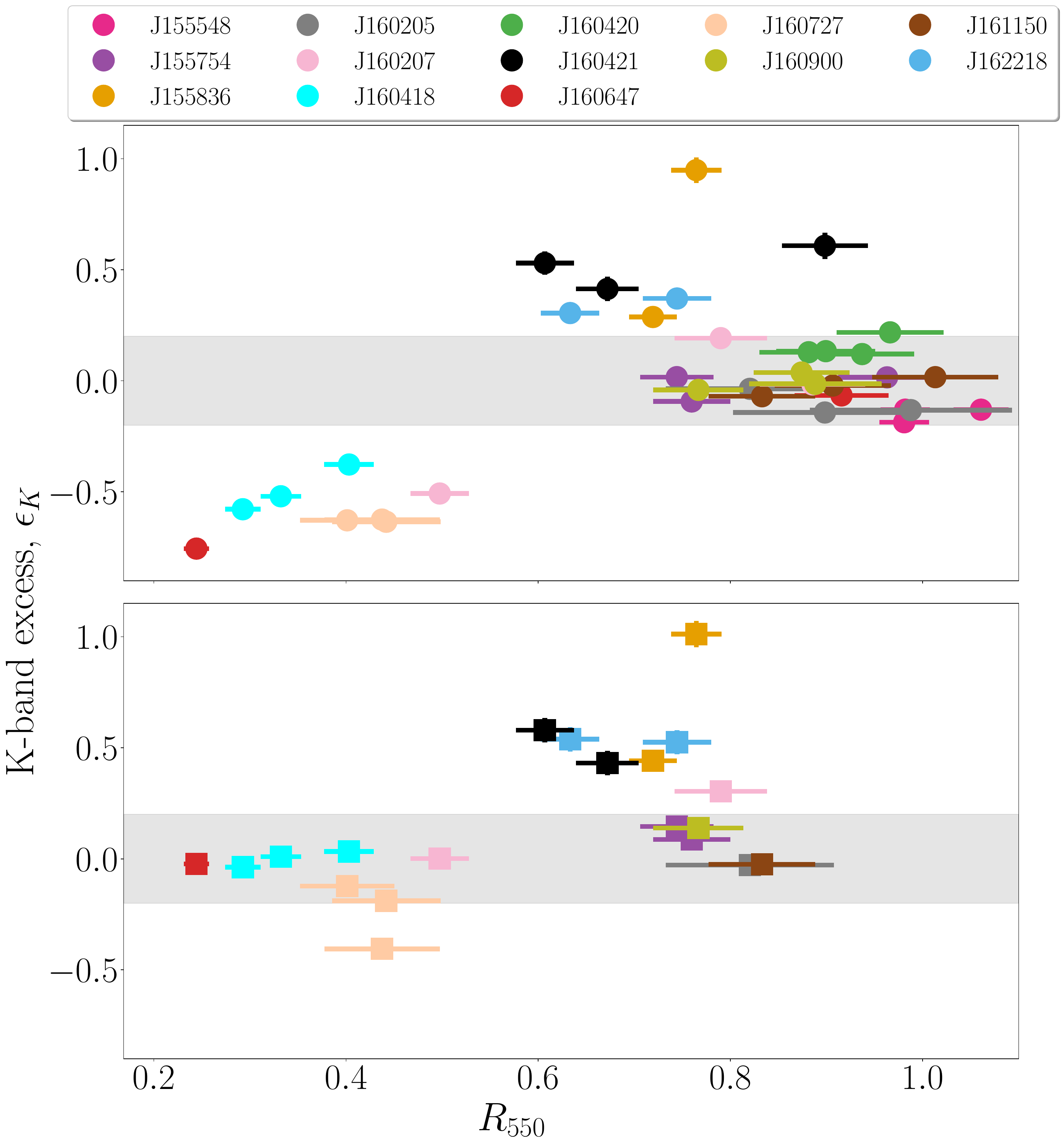}
    \caption{Excess emission at 2.1\,$\mu m$, $\epsilon_K$ shown as a function of \Rvis. The top panel plots the ratio between the observed flux and the intrinsic photospheric one ($\epsilon_K=\frac{F_{obs,K}}{F_{ph,K}}-1$). The bottom panel shows the ratio between the observed flux corrected for the extinction derived at optical wavelengths and the photospheric flux ($\epsilon_K=\frac{F_{obs,K}e^{\tau}}{F_{ph,K}}-1$). The grey area covers a $\pm$20\% region around $\epsilon_K$ = 0. Vertical error bars on $\epsilon_K$ are smaller than the size of the points.}
    \label{fig: nir_excess}
\end{figure}

So far, our analysis has been focussed on the visual range of wavelengths, where the observed flux variations are not affected by the innermost disc hot dust emission. However, an analysis of the NIR excess can potentially shed further light on the dimming events and in particular on the location of the obscuring dust. 

In order to quantify the NIR emission and how it varies among dips, the excess emission was calculated for all objects for which a photospheric level was determined. We define $\epsilon_K$ to represent the amount of excess flux ($F_{obs}, F_{phot}$) at a noise-free region in the K-band of their SEDs (2120-2150 nm). This is calculated as 

\begin{equation}\label{eq: epsilon_k}
    \epsilon_K = \frac{F_{obs}}{F_{phot}} - 1 .
\end{equation}

Discs containing little to no NIR emission in the K-band will therefore have $\epsilon_K \sim 0$, while strong discs will have values $\ge$ 0.2. 

Figure\,\ref{fig: nir_excess} (top panel) shows the computed $\epsilon_K$ values for all the dimming events as a function of \Rvis. At $\epsilon_K$=0, a conservative uncertainty of $\pm$20\% was taken into account (grey area in fig.\,\ref{fig: nir_excess}). All other errors were derived from the RMS of the continuum spectra. As shown in the top panel, most of our measurements are located at $0.0\le\epsilon_K \le 0.2$, and \Rvis $\ge$ 0.75 (i.e. \tauvis $\ge$ 0.28), indicating very little excess emission at NIR wavelengths, and none to little difference between the photospheric flux and the single epoch measurement (i.e. none or small dimming events). This is in agreement with the SEDs of these objects (see fig.\,\ref{fig: master_sed}).  Only seven epochs show $\epsilon_K>$0.2, suggesting stronger disc emission. This is again in agreement with the SEDs of these objects (namely J155836, J162218, and J160421) as shown in fig.\,\ref{fig: master_sed}. Interestingly, eight epochs from four different objects (J160647, J160418, J160727, and J160207) show  $\epsilon_K<0$. This would suggest the unlikely scenario in which the photospheric flux contribution is higher than the observed total flux.

Based on these results, and in order to get hints on the location of the obscuring dust, the excess emission was recalculated by correcting the observed flux by the obscuring dust (local extinction). This was only done for those objects showing dimming events caused by obscuring dust, i.e. with R$_{550}<$0.85 as explained in sect.\,\ref{sec: dust}. Correcting $\epsilon_K$ for the local extinction is equivalent to assuming that the obscuring dust is occulting both the photosphere as well as the hot dust disc emission (assumed to be constant for simplicity). The local extinction was estimated by extrapolating the optical depths of the dust grain distribution (see Sect.\,\ref{sec: grainprop}) to the NIR region. The observed flux was then multiplied by $e^{\tau_\lambda}$. The results are shown in the bottom panel of fig.\,\ref{fig: nir_excess}.

As expected, most of the dimming events have now higher $\epsilon_K$ values after correcting for the local NIR local extinction with only one epoch remaining with $\epsilon_K<0$. It should be noted, however, that for this object (J160727) no multi-epoch Gaia data were available and thus the estimated photospheric level is highly uncertain (see sect.\,\ref{sec: photosphere}). After correction, five epochs show $\epsilon_K\sim0$, while seven dips now show $\epsilon_K>0.3$ indicating a stronger infrared excess emission. Fig.\,\ref{fig: nir_excess}, the dimming events with R$_{550}<$0.6 (i.e. higher optical depths) are those showing the greatest changes in $\epsilon_K$, moving from $\epsilon_K<0$ to $\epsilon_K\sim0$ (with the exception of one epoch as discussed above). In these cases, it is possible that both the photosphere and the NIR disc emission are obscured by the same dust causing the dimming event. Although any conclusions on the location of the obscuring dust are difficult to make given the generally low $\epsilon_K$ values. Interestingly, two objects (namely J155836 and J160207) show epochs with large $\epsilon_K$ changes of 0.57 and 0.30. These variations imply a degree of intrinsic K-band variability. Although due to the small statistics it is difficult to draw further conclusions and ascertain whether these changes are related to those seen in the optical. Similarly, it is also difficult to shed further insights onto the location of the obscuring dust for the remaining dimming events as $\epsilon_K$ variations are likely due to a combination of several factors such as the wavelength dependence of the optical depth, and the strength of the NIR excess.

\section{Accretion variability} \label{sec: accretion}

\begin{figure}
    \centering
    \includegraphics[width=\linewidth]{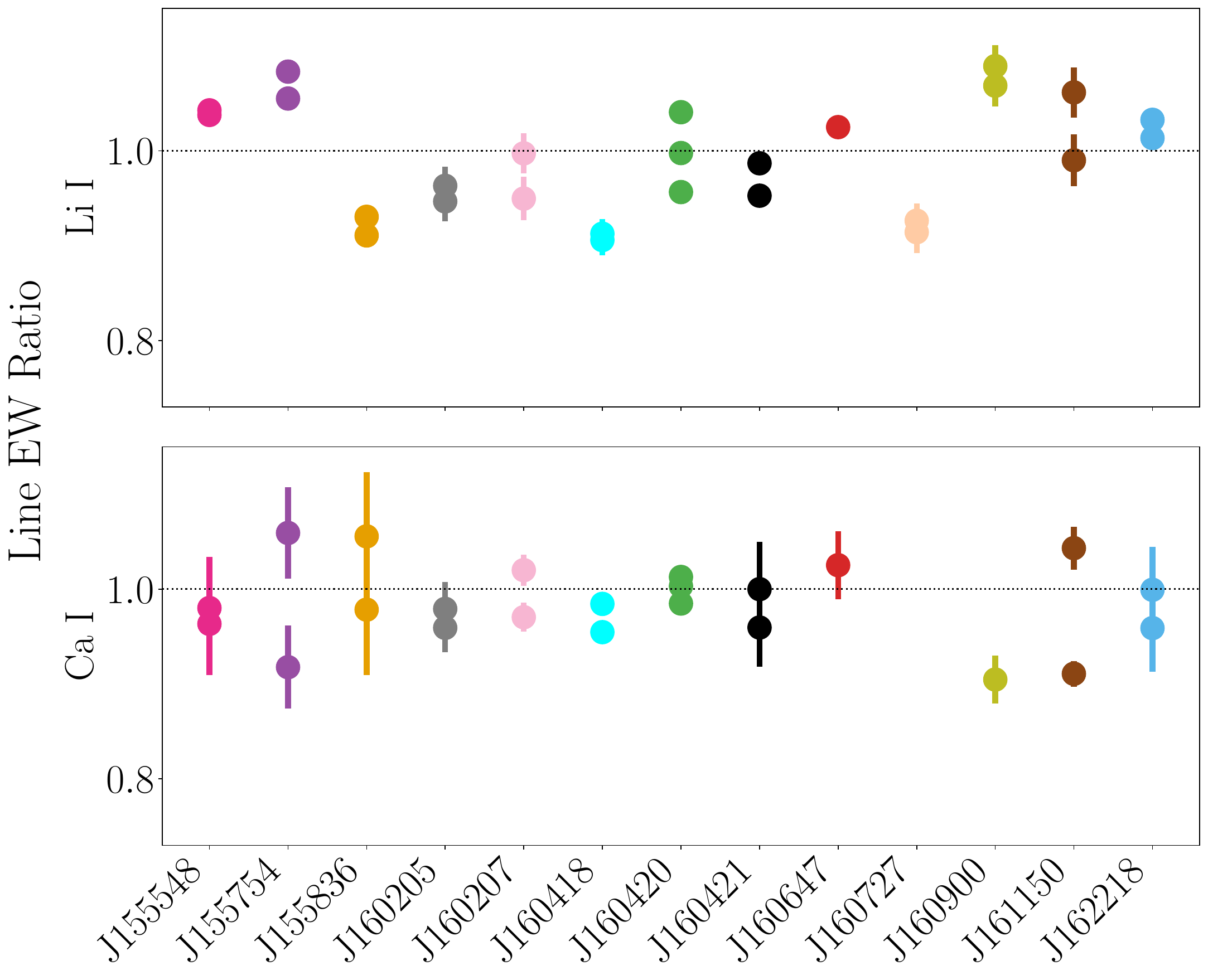}
    \caption{Equivalent width (EW) ratios between epochs of the \LiI\,670.78\,nm line (top panel) and \CaI\,422.67\,nm (bottom panel) photospheric absorption lines. Ratios are calculated with respect to the brightest epoch. Results show that in all cases EW changes are within 10\%. Dotted black lines are at ratios equal to 1, signifying no change in the EW. }
    \label{fig: line_veiling}
\end{figure}

YSOs are also known to undergo short term changes in their levels of accretion, adding a continuum emission to the photospheric one at all wavelengths. The result is a general increase in the observed continuum emission, stronger in the UV, but also clearly detected in the visual, where it causes the photospheric absorption lines to look shallower (veiling). The variable level of local extinction in dippers complicates the typical approach of measuring accretion properties in T Tauri stars from the UV continuum excess \citep{manara13_fitter, manara20, Claes2024}. Measurement of the properties such as the accretion luminosity are therefore beyond the scope of this work. However, to investigate if the observed variability can be explained by changing emission levels, the ratio of the observed Equivalent Widths (EWs) of the \LiI\,(6707.8\,\AA ) and \CaI\,(4226.7\,\AA) lines with respect to each object's brightest epoch was calculated (see Fig.\,\ref{fig: line_veiling}). These lines are well-known proxies of veiling variations. In particular, the \LiI\,6707.8\,\AA\ line is minimally affected by accretion line emission \citep{Campbell-White2023}, while the \CaI\ is a diagnostic of veiling in the blue part of the spectrum \citep{Herczeg2008}. If the photospheric line does not vary, the ratio of the EW between the two spectra should be proportional to the change of the continuum flux at the wavelength of the line. We find in all cases that in general the variations of the EW of both the \LiI\ and \CaI\ lines are <10\%. Therefore, our sample show little evidence to suggest any of the observed variability can be explained by sudden accretion-linked emission changes leading to line veiling.

\section{Discussion}
\label{sec: discussion}

In this work, we presented the most complete sample of irregular dippers in the Upper Scorpius star-forming region to date. In our analysis, we employed a novel approach which consisted of deriving the spectral type of our sources using a sample of Class III templates, and retrieving their quiescent photospheric level from multi-band Gaia photometric observations. This method allowed us to derive the intrinsic photospheric spectrum of 13 out of 16 objects. Unlike previous studies where the study of the dimming events is based on continuous high-cadence photometric light curves \citep{Cody2014, Hedges2018, Roggero2021, Capistrant2022}, here the relative variability is measured with respect to the retrieved photospheric level. This approach therefore relies upon the accuracy in reproducing the true stellar photosphere. We have shown that we are able to reproduce the Gaia observations with a high degree of precision using class III templates (see fig.\,\ref{fig: gaia_magdiff}). As already mentioned (sect.\,\ref{sec: photosphere}) there is an uncertainty of less than two subclasses on the spectral type determination. However, we have found that this difference does not significantly change the derived values of $R_\lambda, \tau_\lambda, \beta$ whose uncertainties are dominated by the noise of the spectra. A side product of this procedure is that the derivation of the interstellar extinction was possible. Accurately identifying the true photosphere, spectral type, and $A_V$ is critical for the studies of such variable objects, as the dimming events can lead to inaccurate measurements of key physical properties such as stellar and accretion luminosities, dust masses, etc. We have shown, for example, that there can be discrepancies between the spectral type (up to three subclasses) and $A_V$ (up to 0.8 mag) measured in this study and those found in literature (\cref{tab: template_results}).

X-Shooter observations allowed us to probe variations in the continuum spectrum from $\sim$400\,nm up to $\sim$2.4\,\micron\, and simultaneously look for accretion variability. Our study showed that our objects show very little variation in their line veiling and therefore the observed photometric changes are not due to sudden accretion variability shocking the stellar surface. Furthermore our analysis concluded that cold spots are also unlikely to reduce the photospheric flux by more that $\sim$15\%. Therefore, dimming events above this limit are believed be due to local dust obscuration, consistent with a number previous studies of irregular dippers \citep{Bouvier1999, Alencar2010a, McGinnis2015, Ansdell2016, Bodman2017,Schneider2018, Sitko2023}.

In this regard, 11 out of 13 objects show an observed flux lower than that of the photosphere at optical wavelengths, indicating the sources were undergoing a dip. The magnitude of the dips varied between sources and observed epochs, ranging from a few percent to $\sim$75\% of the photospheric flux at 550\,nm. While the spread in the degree of variability is common to other studies \citep[e.g.][]{Morales-Calder2011, Ansdell2016, Hedges2018, Alencar2010a, Roggero2021, Venuti2021}, the deepest dips seen in our sample exceed previously recorded ones, which typically showed flux variations of up to 50\%. In particular, our sample show six dips across three targets (namely J160727, J160418 and J160647) that show dips in excess of 50\% of the quiescent photospheric flux level at 550\,nm.  Interestingly, our sample does not show any correlation between the disc inclination and spectral type with the strength of the dimming events. One might expect, for example, that deeper dips occur in systems that are more highly inclined as the line of sight intersects more closely the surface of the disc \citep{Bouvier1999,Lakeland2022}, or that they correlate with higher optical depths. The lack of evidence for such correlations highlights the unpredictability of these objects.

The analysis of the dependence of the dip optical depths on wavelength shows a variety of behaviours, ranging from very flat (low $\beta$ values) to rather steep (high $\beta$) in the optical wavelength range (400-900 nm). 
Under the interpretation that the dimming events are due to obscuring dust, our results show a very dynamical inner disc region. They can have changing dust properties for different dimming events belonging to the same source. Under the assumption of a typical MRN power-law size distribution ($n(a) \propto a^{-q}$) composed of solid astrosilicate grains, our results show that in most cases a top heavy dust grain size distribution is required. In all cases grain sizes larger than those expected in the ISM, are required in order to reproduce the corresponding optical depth wavelength dependence in the 400\,nm--900\,nm range. There is a large variation in a$_{max}$, with no correlation with the depth of the dip. In some cases, our models can only provide lower limits for the maximum grain size with a$_{max}\gtrsim$10-20\,\micron, and in others we require grain growth for $a_{min}$ (up to tenths \micron). Our analysis also shows that in some dips (e.g. J155836) scattering likely plays a significant role, as in several cases the optical depth dependence on wavelength cannot be reproduced by absorption only (see \cref{fig: dustmaps}). While our models do test the extreme case of multiple scatterings (pure absorption), we focus on understanding the contribution to the extinction from single scattering events only. We find that in many cases the albedo of the local extinction is different from that of typical ISM conditions, with an increased contribution of scattering. Under the assumption of classical extinction ($k_{abs} + k_{sca}$), it is the scattering that dominates above the absorption allowing for the unusual $\tau_\lambda$ curves to be possible (see \cref{fig: dustmaps}, $\tau_\lambda$ plot). This arises from the transition of Rayleigh to Mie scattering regimes. This increased scattering effect is a result of the changing grain sizes and the fact that the dips are typically optically thin, instead of any geometrical effects. The scattering contribution is important to consider when observing such targets, and has been previously linked with the dipping mechanisms \citep{Bouvier1999, Bouvier2003, Dodin2019}. Interestingly, in these cases the dips are linked to very narrow grain size distributions with both a$_{min}$ and a$_{max}$ of the order of tenths of microns. Given the typical timescale of the dips, the obscuring dust is understood to be coming from the inner disc regions \citep[<<1 au;][]{Morales-Calder2011}. The fact that dust grains of sizes up to tens of $\mu m$ exist in these regions suggests some level of grain growth has occurred \citep{Nagel2020,Sitko2023}.

Regarding the NIR emission at 2\,\micron, by assuming the same dust model as those derived at optical wavelengths and propagating into the NIR we were able to give some constraints on the location of the obscuring dust. Although the short time scale variability of dippers is generally linked to the closest-in location of the obscuring dust \citep[e.g.][]{Morales-Calder2011}, here we were able show that in some cases it may be possible that both the photosphere and region emitting the 2$\mu m$ continuum emission are obscured by the same dust. In most objects, the extinction-corrected 2$\mu m$ continuum emission does not vary between epochs. However, in two cases (J155836, J160207), there remains a difference in excess emission between two dips. This suggests an intrinsically varying NIR emission for these objects. However, in general low NIR excess emissions combined with the range of possible uncertainties make it difficult to draw firm conclusions.

Although similar results regarding the location and spread in dust grain size distributions, including for some objects in our sample, have also been reported in literature \citep[see e.g.][]{Bouvier1999, Bouvier2003, Schneider2018, Sitko2023}, in this study, we show the presence of relatively quick changes in the grain size distribution in the inner disc. With time-scales ranging from a few weeks to several months, the dust grain size distributions can change from epoch to epoch. The distributions in some cases are top-heavy and others they have narrow sub-micron dust grain distributions dominated by scattering. The changes are in general too quick to be explained by simple radial drag of different dust grain populations migrating from outer to inner regions of the disc. This points towards some mechanism capable of accessing various grain sizes from differing depths in a vertically stratified disc structure.
Recent ALMA and SPHERE observations of protoplanetary discs show indeed the presence of different grain size populations at similar spatial scales \citep[e.g.][]{Andrews2018,Garufi2024}. For instance, SPHERE scattered light images detect stellar radiation that is scattered off small dust grains at the disc surface, whereas at the same radii, ALMA dust continuum emission reveals the presence of much larger dust grains at deeper vertical depths \citep[e.g. IM\,Lup][]{Andrews2018, Avenhaus2018}. 

One mechanism capable of lifting such dust grains to significant heights above the disc plane to obscure the star, and sometimes also the region emitting at 2\,\micron , independent of the disc inclination, is the so-called dusty disc wind. In recent years several studies have investigated the entrainment of dust grains in MHD and photoevaporative winds \citep{Owen2011, Miyake2016, Giacalone2019, Booth2021}. These models predict different regions from which these winds will be launched as well as different sizes of dust grains that can be lifted at those radii. In general, models predict a maximum grain size that can be lifted from a certain distance from the central source. Some of these models also include a vertical dust grain size stratification with grain sizes at 1\,au ranging from sub-micron to one hundred micron \citep[see e.g.][]{Miyake2016}. For a vertical cut in the disc, at a certain distance from the source, the size of the grains that can be lifted depends on the mass flux of the disc wind \citep{Miyake2016, Giacalone2019}. Therefore, variations in the mass flux of the disc wind might explain the changes observed in the grain size distributions needed to reproduce the observed wavelength dependence of the optical depth in the optical. Dusty disc winds have been previously used to explain the dimming events in RW\,Aur \citep{Dodin2019} as well as scattered light observations of the disc surface of IM\,Lup \citep{Avenhaus2018}, and in general linked with YSO variability \citep{Ellerbroek2014,Lakeland2022,Grinin2023,Sitko2023}.

Alternatively, dust entrainment along magnetospheric accretion columns could also explain the presence of the obscuring dust at high scale heights from the disc surface. Stable and unstable accretion columns have been shown to be able to lift dust grains above the disc surface following the magnetic field lines \citep{Romanova2003,Nagel2024}. The evaporation of the dust grains approaching the central star may explain the different levels of optical depths encountered. While this configuration tends to favour occultation of systems at high inclinations, varying dust survival along the accretion columns might be able to explain lower inclinations down to 45\degr\ \citep{Nagel2024}.
In this context, unstable accretion columns would be particularity favourable for irregular dippers given their short timescales and lack of periodicity, and was believed to be the case for four systems (J160418, J160205, J160727 and J160900) in this sample \citep{Bodman2017}. Stellar occultation by disc warps in the inner regions, arising from such accretion events, is also a possibility \citep{Romanova2013,Sicilia-Aguilar2020, McGinnis2015}. Instabilities arising from such mechanisms have also been shown to affect the dust grain growth \citep{Aly2024}. Given that a number of the objects in this sample have been linked to this accretion driven variability, we cannot rule it out as a possible dip driving mechanism.

\section{Summary and conclusions}\label{sec: conclusions}

In this paper we presented the results of a programme aimed at studying the properties of a number of irregular dippers in the Upper Scorpius star-forming region. The sample includes 16 objects, each of them was observed with X-Shooter two to four times, with time intervals ranging from days to years, for a total of 46 spectra. 
Our main results are summarised below:

\begin{itemize}
    \item By combining multi-epoch Gaia photometry with the spectral type derived from Class III templates, we find evidence that for 11 out of 13 objects the observed emission was lower than that of the photosphere, that is the star was undergoing a dip.
    \item The magnitude of the dips varies between epochs even for the same star ranging from few percent to up to $\sim$75\% of the photospheric flux at 550\,nm.
    \item Assuming that the dips are caused by dust obscuration, we find that in most cases the grains causing the dips favour a top-heavy size distribution (i.e. most of the dust mass is in grains significantly bigger than those in the ISM). We find in some cases lower limits on the maximum grain size to be tens of microns. However, for a few cases, scattering also plays a significant role. In this case, a narrow grain distribution of sub-micron dust grains is needed.
    \item The properties of the obscuring dust changes from dip to dip and from source to source, with no evident relationship between the properties of the dip and stellar properties such as spectral type or disc inclination.
    \item By analysing the continuum emission at 2\,\micron\ it was shown that most of our objects do not show significant NIR excess.
    \item By assuming the same dust model as at optical wavelengths, the extinction at 2\,\micron\ was estimated to constrain the location of the obscuring dust. Although a detailed modelling of the dust properties is beyond the scope of this paper, under this simple assumption it was possible to infer that in a few sources the same dust may be obscuring both the source and the 2\,\micron\ emitting region of the disc. A few sources also show hints of 2\,\micron\ continuum variability. 
\end{itemize}

In general, the picture that comes out of our results is that of very variable phenomena, not just in the depth of the dips but also in terms of the properties of the dust that cause them. This is true not only between different stars, but also for the same star, that can have dips due to grains of different properties, with no obvious trend that links that to the magnitude of the dip. These changes in the dust properties of different dips also point towards a highly dynamic dust growth and/or transport process in the inner disc. To further answer the questions on dipper variability, simultaneous spectroscopic and photometric observations covering full dimming events would be highly beneficial. Additionally, scattered light observations could inform of the presence of the dust substructures we believe to causing the dips. 

\section*{Data availability}
This project was based on observations collected at the European Southern Observatory under ESO programmes 0105.C-0513(A), 097.C-0378(A), 0101.C-
0866(A), and 0103.C-0887(B). Reduced spectra can be found on the ESO Science Archive Facility with DOI: https://doi.eso.org/10.18727/archive/71

\begin{acknowledgements}
   We thank the anonymous referee for their careful consideration their comments that helped improve the clarity of our work.

    A.E. acknowledges funding from Taighde Éireann – Research Ireland under Grant number GOIPG/2023/4396 and the UCD Physics Scholarship in Research and Teaching (SIRAT).
    
    CFM is funded by the European Union (ERC, WANDA, 101039452). Views and opinions expressed are however those of the author(s) only and do not necessarily reflect those of the European Union or the European Research Council Executive Agency. Neither the European Union nor the granting authority can be held responsible for them. A.E. acknowledges support from the ESO Early-Career Scientific Visitor Programme when part of this work was carried out.

    MB has received funding from the European Research Council (ERC) under the European Union’s Horizon 2020 research and innovation programme (PROTOPLANETS, grant agreement No. 101002188).

    This work has made use of data from the European Space Agency (ESA) mission {\it Gaia} (\url{https://www.cosmos.esa.int/gaia}), processed by the {\it Gaia} Data Processing and Analysis Consortium (DPAC, \url{https://www.cosmos.esa.int/web/gaia/dpac/consortium}). Funding for the DPAC
    has been provided by national institutions, in particular the institutions participating in the {\it Gaia} Multilateral Agreement. 
    
    This research has made use of the NASA Exoplanet Archive, which is operated by the California Institute of Technology, under contract with the National Aeronautics and Space Administration under the Exoplanet Exploration programme.
\end{acknowledgements}

\bibliographystyle{aa}  
\bibliography{ref}

\newpage
\onecolumn
\begin{appendix}

\section{Spectral energy distributions} \label{sec: AppA_data}
\begin{figure*}[h!]
    \centering
    \includegraphics[width=0.8\linewidth]{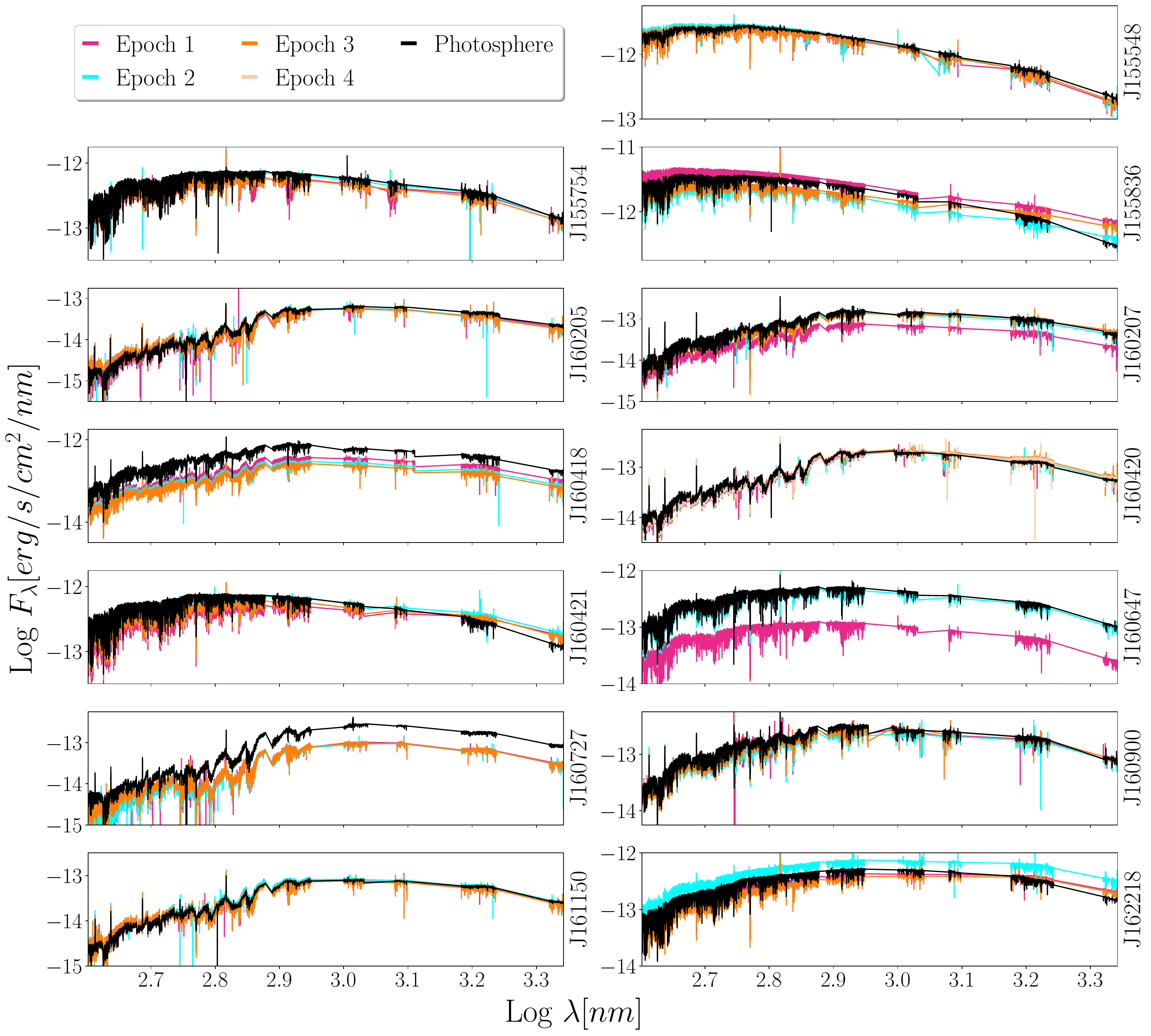}
    \caption{Observed spectra for each epoch are plotted alongside the  ISM-reddened photosphere (where applicable).}
    \label{fig: master_sed}
\end{figure*}

\Cref{fig: master_sed}
 shows the observed spectra for all the targets where the photospheric spectrum could be derived (see sect.~\ref{sec: photosphere}).

\begin{figure}[h!]
    \centering
    \includegraphics[width=0.7\linewidth]{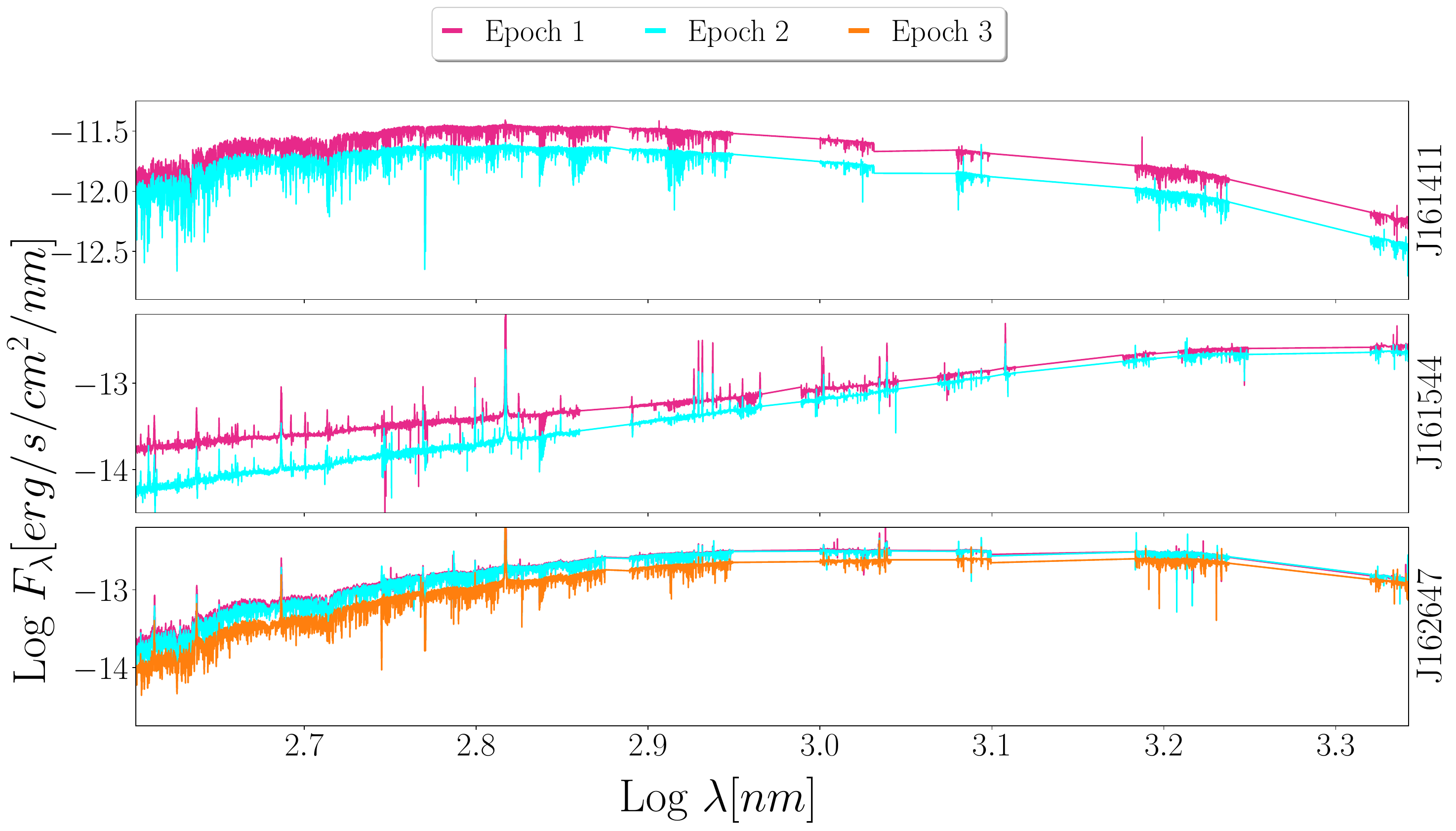}
    \caption{Observed spectra for objects for which a photosphere could not be derived.}
    \label{fig: oddobjects_sed}
\end{figure}

\newpage

\section {Gaia photometry} \label{sec: AppB_Gaia}
\begin{figure*}[h!]
    \centering
    \includegraphics[width=0.8\linewidth]{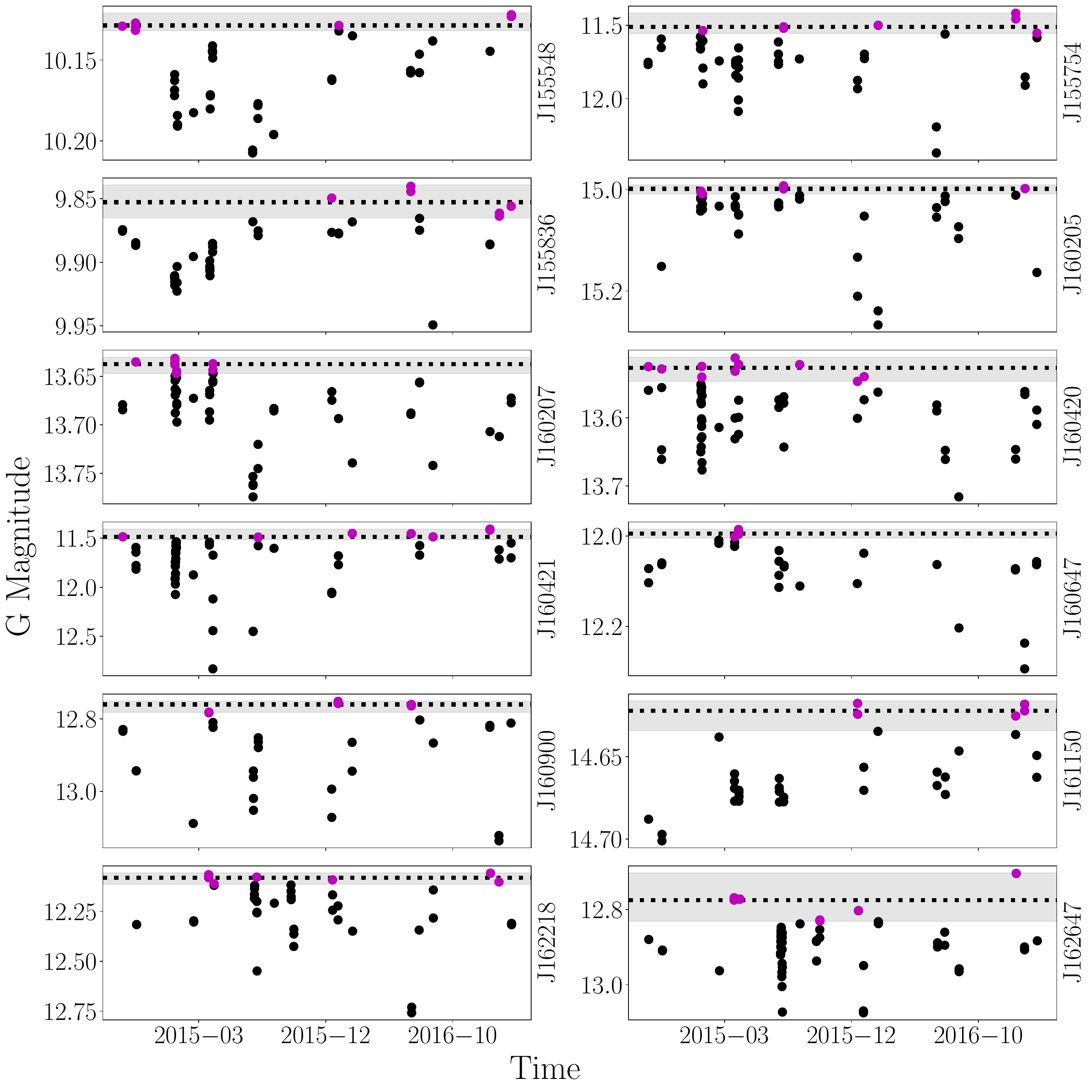}
    \caption{G-band magnitudes observed by the Gaia Variability programme \citep{Gaia2016, GaiaDR3} for all the objects included in the Gaia dataset.  In each panel, the observed magnitudes are shown by dots. The  shaded region defines the range between the minimum and 85\% of it (i.e. the range of the brightest values). The dotted line shows the median of the points included in this range (in magenta), and is assumed to be the G-band magnitude of the stellar photosphere.}
    \label{fig: gaia_appendix}
\end{figure*}

We show in this section (fig.~\ref{fig: gaia_appendix} the observed photometric broad-band (G) magnitudes available  from the Gaia Variability programme \citep{Gaia2016, GaiaDR3}. In each panel, the dots show the observed value. The grey area  highlight the range of magnitudes between the maximum and 85\% of it. The dashed line plots the median of the G values included in the region (in magenta), that we assume to be representative of the photosphere of the object.

\begin{figure}[h!]
    \centering
    \includegraphics[width=0.5\linewidth]{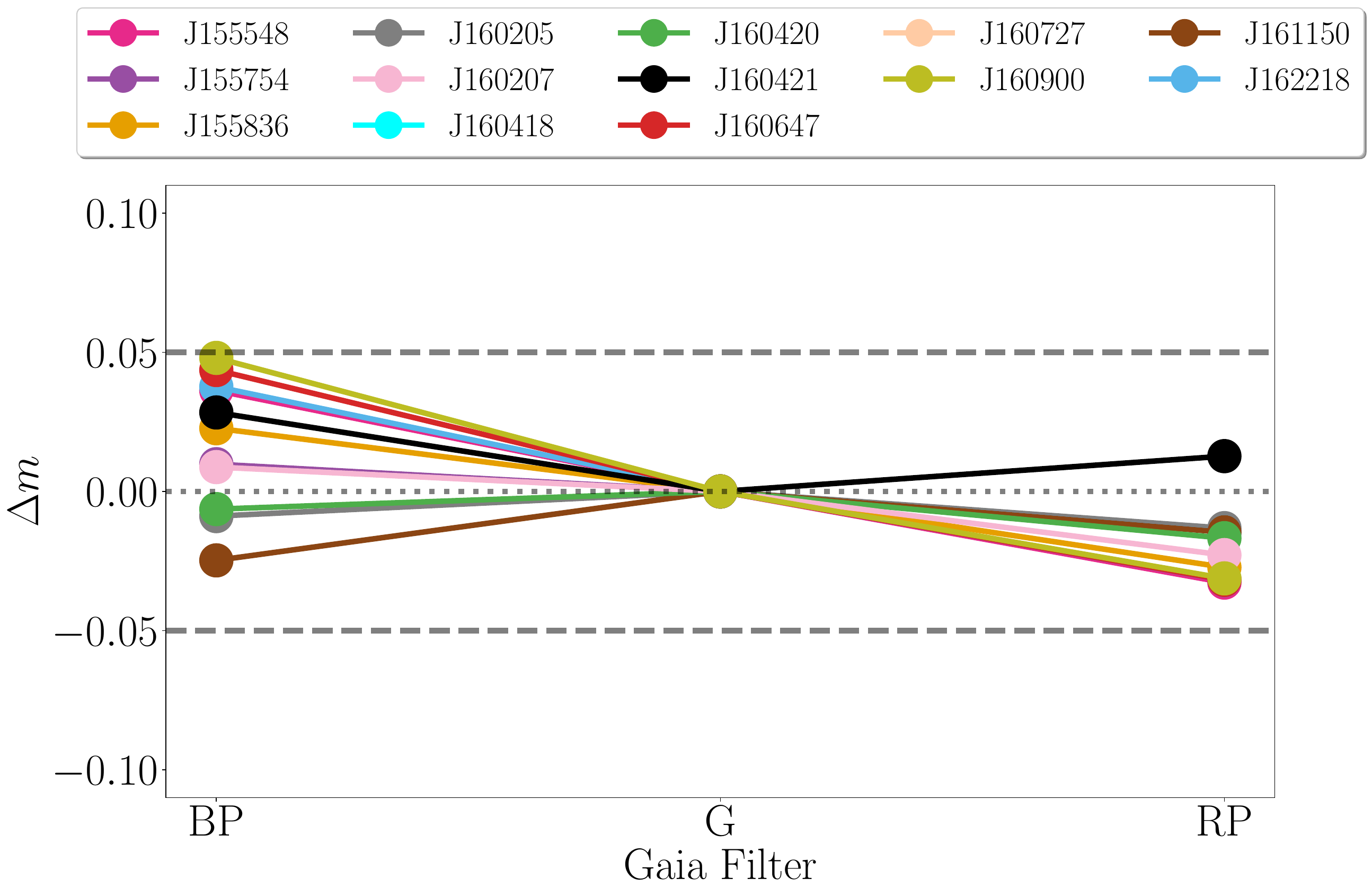} 
    \caption{Difference in magnitude between the synthetic magnitudes of the reddened adopted photosphere and the Gaia observed values for the BP and RP filters.  By definition, G-band difference between the reddened adopted photosphere and the observed magnitude is 0. Dashed lines mark the $\pm$5\% interval around $\Delta m = 0$.} 
    \label{fig: gaia_magdiff}
\end{figure}

The procedure followed to derive the ISM extinction and the photospheric spectrum for each object is described in sect.~\ref{sec: photosphere}.  Fig.~\ref{fig: gaia_magdiff} shows for each object the difference  between the reddened synthetic magnitude of the adopted photosphere  and the observed values. The difference is always smaller than 5\%.

\newpage
\section {Ratios} \label{sec: AppC_ratios}
\begin{figure*}[h!]
    \centering
    \includegraphics[width=0.85\linewidth]{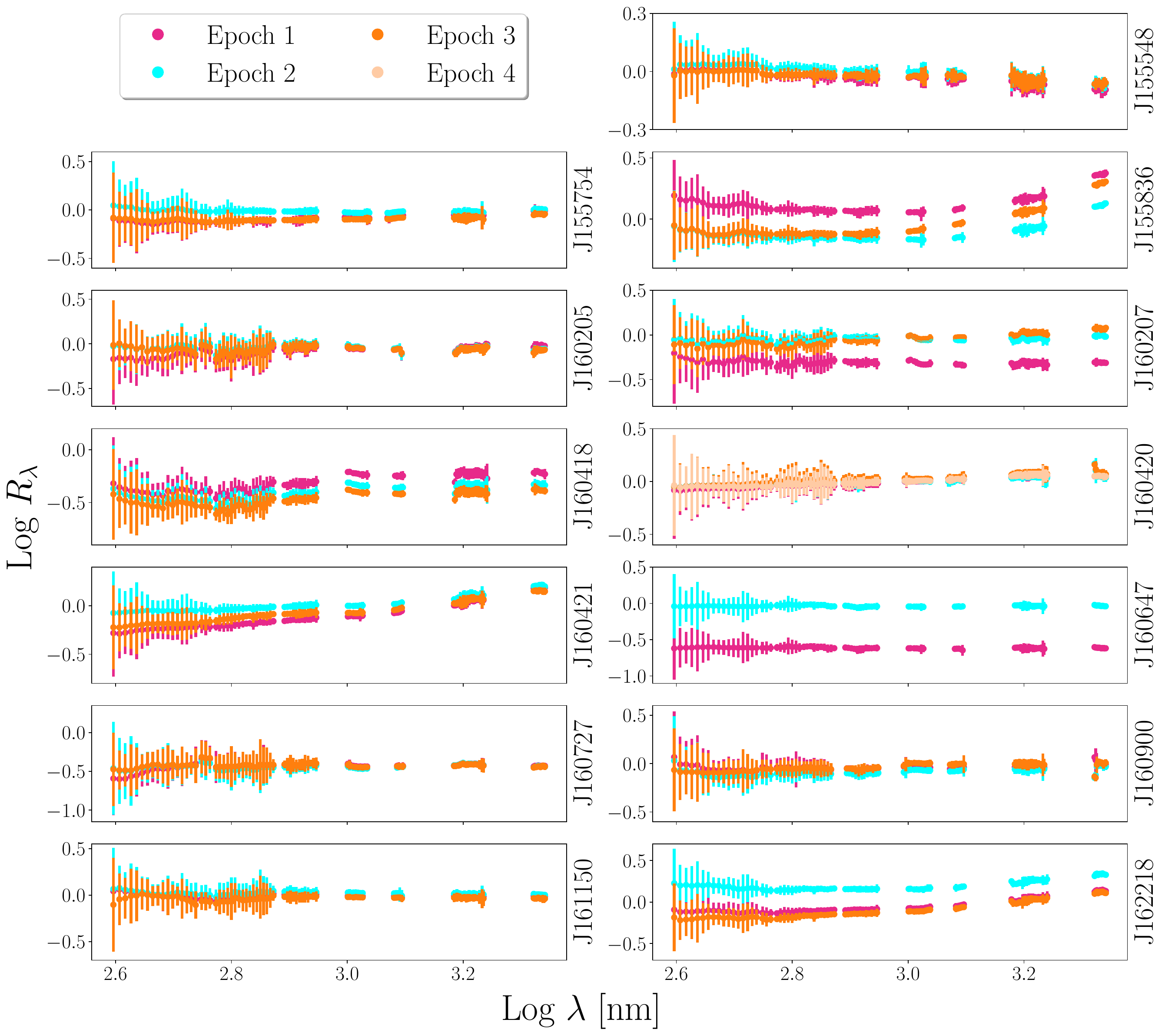}
    \caption{ Ratio R$_\lambda$ of the observed over the ISM-reddened photospheric flux versus  $\lambda$. The ratios are computed from the spectra binned in equal intervals of 95\AA\ and $\sigma$-clipped.}
    \label{fig: gaia_R_masterplot}
\end{figure*}

Fig.~\ref{fig: gaia_R_masterplot} shows the values of the ratio of the observed to the ISM-reddened photospheric fluxes as function of the wavelength.

\newpage
\section{Grain properties} \label{sec: AppD_tau}
\begin{figure*}[!h]
    \centering
    \includegraphics[width=0.8\linewidth]{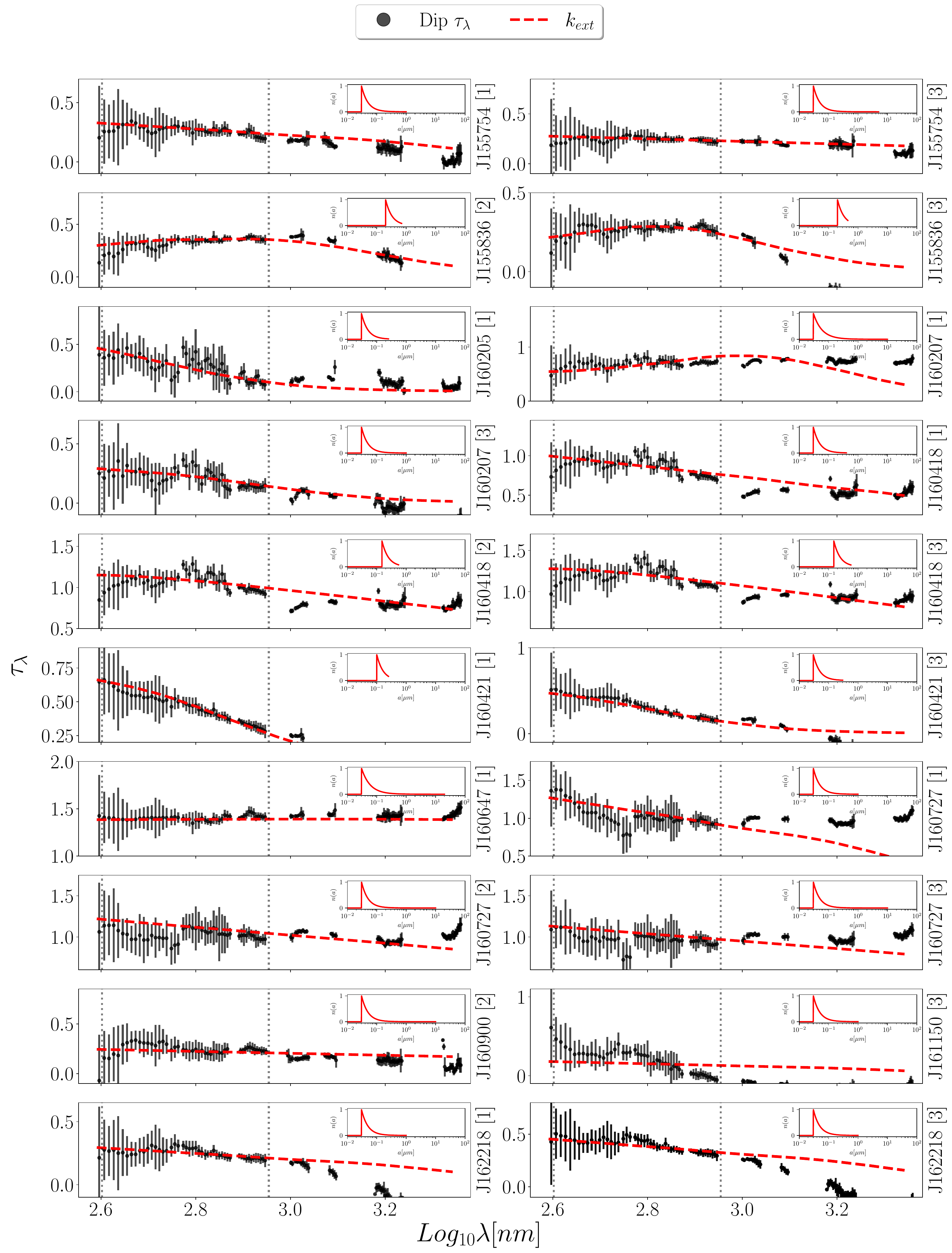}
    \caption{$\tau_\lambda$ as a function of $\lambda$ for all the 20 dips analysed in Sect.\ref{sec: dust}. The name of the star and the epoch number are given on the secondary axis.
    The observed values are shown by black dots with error bars.  The red coloured lines show the results for the grain size distribution that best fit the data in the wavelength interval 400-900 nm for the extinction opacities, each normalised to the observed value at 550 nm. The values of $a_{min}, a_{max}, q$ are given on top of each panel; the corresponding  grain size distribution is displayed in the insert. The vertical grey dotted lines represent the boundaries of $\lambda = 400, 900$ nm. Three representative cases are also shown in the main paper.}
    \label{fig: master_tau}
\end{figure*}

Fig.~\ref{fig: master_tau}  shows the wavelength-dependence of the optical depth for the 20 dips studied in Sect.~\ref{sec: grainprop}.

\newpage
\section{Individual dippers} \label{sec: AppC_cases}

Of the objects in this study, two are particularly noteworthy for being well studied in the context of YSO variability, J155836 and J160421. 

\subsection{J155836}
J155836, often referred to as HD~143006 in the literature, is a  K0 T Tauri star \citep{Manara2020} (M$_\ast \sim 1.5$ M$_\odot$) with shallow, irregular photometric dips \citep{Ansdell2018}. Its outer disc is highly structured, with a series of rings and gaps, and a bright arc exterior to the rings apparent at high angular resolution in sub-millimetre images \citep{Andrews2018, PerezLaura2018}. In the infrared scattered light, the disc appears to show a large-scale asymmetry, with the eastern side of the disc brighter than the western side \citep{Benisty2018}, indicating a shadow cast by a potentially misaligned inner disc region, inaccessible to direct imaging observations. Interestingly, the innermost $\sim$7 au of the disc appear depleted of large dust grains (emitting in the sub-mm) \citep{Jennings2022}, and the ring beyond the cavity could be misaligned by $\sim$40$^\circ$ with respect to the outer disc \citep{PerezLaura2018} possibly due to a strongly inclined binary companion within the cavity \citep{Ballabio2021}. Nevertheless, the cavity is not fully depleted of material as the star is accreting \citep{Manara2020} and the emission of small micron-sized dust at the dust sublimation edge is spatially resolved with the Very Large Telescope Interferometer \citep{Lazareff2017}.  

\subsection{J160421}
In contrast, J160421, a K3 star, shows  negligible accretion \citep{Manara2020}. It hosts a nearly face-on transition disc with a large depleted cavity while being one of the most massive discs in Upper Sco \citep{Barenfeld2017}. Its outer disc resolves into a single ring \citep{Mayama2012,Mayama2018} that has clear indication of being shadowed by a misaligned inner disc. Unlike the case of J155836, the shadows appear azimuthally narrow \citep{Mayama2012, Pinilla2018} and highly variable in both their location and morphology within very short timescales of a day. Some of the properties of the shadows can be explained with a broken circumbinary disc, and a precessing, strongly misaligned, inner disc \citep{Nealon2020}. Additional support for the presence of a misaligned inner disc responsible for the shadows are found in the warped kinematics of the gas within the cavity  and in the shadow counterparts in the sub-mm \citep{Mayama2018}. \citet{Davies2019} analysed projected rotational velocities and determined that the stellar rotation axis inclination angle strongly differs from the one of the outer disc, indicating that the inner disc and the star are well aligned. This system was qualified as a dipper from its K2 light curve, and shows aperiodic dimming events of $\sim$1 mag in depth, superposed onto a $\sim$5.1 days periodic signal \citep{Ansdell2016,Ansdell2018}. Time-resolved multi-wavelength photometric studies recover the $\sim$5 days quasi-periodicity but show variability in the eclipse depths and frequency,  indicating an irregularly shaped inner disc, whose mass is completely accreted (and then replenished) in days to weeks timescales \citep{Sicilia-Aguilar2020}.

\newpage
\section{Log of the observations}\label{sec: AppD_observations}

\begin{table*}[h!]
\centering
\caption{Night log of observations.}\label{tab: logobs}
\begin{tabular}{c c c c c c}
\hline
\hline
2MASS ID          & Date of Observation          & Exp. Time & \multicolumn{3}{c}{Slit Width ['']} \\ \cline{4-6}
                  &   {[}UT{]}                   &  [Nexp x (s)]         & UVB    & VIS    & NIR    \\
\hline
                  &                              &           &        &        &        \\
J15554883-2512240 & 2021-01-16T08:35:25.742      & 4x120     & 1.0x11 & 0.4x11 & 0.4x11 \\
                  & 2021-06-09T05:01:44.708      & 4x100     & 1.0x11 & 0.4x11 & 0.4x11 \\
                  & 2021-07-07T05:20:01.503      & 4x120     & 1.0x11 & 0.4x11 & 0.4x11 \\
J15575444-2450424 & 2021-05-04T01:15:14.066      & 4x200     & 1.0x11 & 0.4x11 & 0.4x11 \\
                  & 2021-07-15T02:47:11.784      & 4x200     & 1.0x11 & 0.4x11 & 0.4x11 \\
                  & 2021-08-05T02:05:18.219      & 4x200     & 1.0x11 & 0.4x11 & 0.4x11 \\
J15583692-2257153 & 2016-07-25T03:37:53.376      & 4x120     & 0.5x11 & 0.4x11 & 0.4x11 \\
                  & 2018-05-20T23:59:35.513      & 4x150     & 0.5x11 & 0.4x11 & 0.4x11 \\
                  & 2019-06-24T01:39:37.872      & 4x150     & 1.0x11 & 0.4x11 & 0.4x11 \\
J16020517-2331070 & 2021-08-22T01:59:54.065      & 4x420     & 1.0x11 & 0.4x11 & 0.4x11 \\
                  & 2021-09-15T00:34:41.519      & 4x420     & 1.0x11 & 0.4x11 & 0.4x11 \\
                  & 2021-09-19T23:46:54.111      & 4x420     & 1.0x11 & 0.4x11 & 0.4x11 \\
J16020757-2257467 & 2018-05-20T00:39:53.156      & 4x195     & 1.0x11 & 0.4x11 & 0.4x11 \\
                  & 2018-05-21T08:02:53.116      & 4x150     & 1.0x11 & 0.4x11 & 0.4x11 \\
                  & 2019-06-07T04:25:40.574      & 4x450     & 1.0x11 & 0.4x11 & 0.4x11 \\
J16041893-2430392 & 2021-05-15T09:16:15.673      & 4x450     & 1.0x11 & 0.4x11 & 0.4x11 \\
                  & 2021-07-05T06:05:34.303      & 4x450     & 1.0x11 & 0.4x11 & 0.4x11 \\
                  & 2021-07-11T04:19:08.429      & 4x450     & 1.0x11 & 0.4x11 & 0.4x11 \\
J16042097-2130415 & 2021-08-04T00:32:35.772      & 4x430     & 1.0x11 & 0.4x11 & 0.4x11 \\
                  & 2021-08-20T00:30:55.210      & 4x430     & 1.0x11 & 0.4x11 & 0.4x11 \\
                  & 2021-08-22T02:43:35.395      & 4x420     & 1.0x11 & 0.4x11 & 0.4x11 \\
                  & 2021-09-05T00:57:10.754      & 4x420     & 1.0x11 & 0.4x11 & 0.4x11 \\
J16042165-2130284 & 2016-08-18T02:43:53.101      & 4x150     & 0.5x11 & 0.4x11 & 0.4x11 \\
                  & 2018-05-21T08:55:04.624      & 4x120     & 0.5x11 & 0.4x11 & 0.4x11 \\
                  & 2019-06-19T07:04:34.387      & 4x225     & 1.0x11 & 0.4x11 & 0.4x11 \\
J16064794-1841437 & 2021-07-14T03:55:56.378      & 4x300     & 1.0x11 & 0.4x11 & 0.4x11 \\
                  & 2021-08-02T00:13:22.092      & 4x300     & 1.0x11 & 0.4x11 & 0.4x11 \\
J16072747-2059442 & 2021-07-14T03:07:28.062      & 4x450     & 1.0x11 & 0.4x11 & 0.4x11 \\
                  & 2021-08-21T02:42:42.196      & 4x430     & 1.0x11 & 0.4x11 & 0.4x11 \\
                  & 2021-09-01T01:49:05.120      & 4x420     & 1.0x11 & 0.4x11 & 0.4x11 \\
J16090075-1908526 & 2018-05-20T09:34:05.964      & 4x75      & 1.0x11 & 0.4x11 & 0.4x11 \\
                  & 2018-05-21T08:33:53.141      & 4x75      & 1.0x11 & 0.4x11 & 0.4x11 \\
                  & 2019-06-24T01:02:58.791      & 4x330     & 1.0x11 & 0.4x11 & 0.4x11 \\
J16115091-2012098 & 2021-08-07T02:13:17.411      & 4x420     & 1.0x11 & 0.4x11 & 0.4x11 \\
                  & 2021-09-08T00:02:02.638      & 4x420     & 1.0x11 & 0.4x11 & 0.4x11 \\
                  & 2021-09-20T00:32:21.210      & 4x420     & 1.0x11 & 0.4x11 & 0.4x11 \\
J16141107-2305362 & 2021-05-02T06:39:46.823      & 4x220     & 1.0x11 & 0.4x11 & 0.4x11 \\
                  & 2021-07-14T00:30:27.996      & 4x220     & 1.0x11 & 0.4x11 & 0.4x11 \\
J16154416-1921171 & 2021-06-29T06:20:19.140      & 4x450     & 1.0x11 & 0.4x11 & 0.4x11 \\
                  & 2021-08-05T02:33:24.175      & 4x450     & 1.0x11 & 0.4x11 & 0.4x11 \\
J16221852-2321480 & 2021-07-15T03:14:42.806      & 4x300     & 1.0x11 & 0.4x11 & 0.4x11 \\
                  & 2021-07-18T04:48:23.932      & 4x300     & 1.0x11 & 0.4x11 & 0.4x11 \\
                  & 2021-08-07T04:27:16.280      & 4x300     & 1.0x11 & 0.4x11 & 0.4x11 \\
J16264741-2314521 & 2021-06-29T07:06:23.599      & 4x200     & 1.0x11 & 0.4x11 & 0.4x11 \\
                  & 2021-07-18T05:23:18.158      & 4x200     & 1.0x11 & 0.4x11 & 0.4x11 \\
                  & 2021-08-07T02:55:24.473      & 4x200     & 1.0x11 & 0.4x11 & 0.4x11 \\
                  \hline
\end{tabular}
\end{table*}

\end{appendix}
\label{LastPage}
\end{document}